# Amide Proton Transfer (APT) imaging in tumor with a machine learning approach using partially synthetic data.

**Malvika Viswanathan [1,2], Leqi Yin[3], Yashwant Kurmi [1,4], Zhongliang Zu [1,2,4]**

[1]Vanderbilt University Institute of Imaging Science, Vanderbilt University Medical Center, Nashville, US. [2]Department of Biomedical Engineering, Vanderbilt University, Nashville, US. [3]School of Engineering, Vanderbilt University, Nashville, US. [4]Department of Radiology and Radiological Sciences, Vanderbilt University Medical Center, Nashville, US.

**Word counts: Total 5212; Abstract 239.**

Correspondence to:

Zhongliang Zu, Ph.D.

Vanderbilt University Institute of Imaging Science

1161 21st Ave. S, Medical Center North, AAA-3112

Nashville, TN 37232-2310

Email: zhongliang.zu@vumc.org

Phone: 615-875-9815

Fax: 615-322-0734

Grant Sponsor: R21 AR074261, R03 EB029078, R01 EB029443

Running title: Machine learning-based APT imaging

**ABSTRACT**

**Purpose:** Machine learning (ML) has been increasingly used to quantify chemical exchange saturation transfer (CEST) effect. ML models are typically trained using either measured data or fully simulated data. However, training with measured data often lacks sufficient training data, while training with fully simulated data may introduce bias due to limited simulations pools. This study introduces a new platform that combines simulated and measured components to generate partially synthetic CEST data, and to evaluate its feasibility for training ML models to predict amide proton transfer (APT) effect.

**Methods:** Partially synthetic CEST signals were created using an inverse summation of APT effects from simulations and the other components from measurements. Training data were generated by varying APT simulation parameters and applying scaling factors to adjust the measured components, achieving a balance between simulation flexibility and fidelity. First, tissue-mimicking CEST signals along with ground truth information were created using multiple-pool model simulations to validate this method. Second, an ML model was trained individually on partially synthetic data, in vivo data, and fully simulated data, to predict APT effect in rat brains bearing 9L tumors.

**Results:** Experiments on tissue-mimicking data suggest that the ML method using the partially synthetic data is accurate in predicting APT. In vivo experiments suggest that our method provides more accurate and robust prediction than the training using in vivo data and fully synthetic data.

**Conclusion:** Partially synthetic CEST data can address the challenges in conventional ML methods.

**Key words:** Chemical exchange saturation transfer (CEST), amide proton transfer (APT), machine learning, tumor

## INTRODUCTION

Chemical exchange saturation transfer (CEST) is a magnetic resonance imaging (MRI) contrast mechanism that enables the detection of low-concentration solute molecules with exchangeable/coupling protons, which are not visible using conventional MRI techniques (1-5). In CEST imaging, a long radio frequency (RF) pulse is used to saturate the solute protons, and the resulting change in the water signal due to saturation transfer is measured. By plotting the water signal as a function of the RF frequency offset, a CEST Z-spectrum is obtained, allowing the observation of all CEST effects arising from exchangeable or coupling protons. One specific CEST effect of interest is the amide proton transfer (APT) effect, which occurs at 3.5 ppm from water. APT arises from the chemical exchange between amide protons in mobile proteins or peptides and water protons (6,7). APT has gained significant attention as a promising technique for protein imaging and pH imaging. It has demonstrated potential in various diagnostic applications, including tumor detection (8-11), ischemic stroke (12-14), multiple sclerosis (15,16), traumatic brain injury (17), Alzheimer's disease (18-22), and Parkinson's disease (23,24).

Until now, achieving specific APT imaging has been a challenging task due to various confounding factors. APT signals in biological tissues are influenced not only by the APT effect but also by overlapping signals from the direct water saturation (DS) effect, semisolid magnetization transfer (MT) effect, and nearby amine CEST effect. Multiple-pool Lorentzian fit is a commonly used data analysis method to isolate the APT effect from confounding signals. However, its accuracy strongly depends on the fitting model as well as the imaging signal-to-noise ratio (SNR), initial values, boundaries, and experimental conditions. Recently, machine learning methods has emerged as a promising approach for quantifying APT effects, offering improved robustness and processing speed compared to the multiple-pool Lorentzian fit (25-27). Typically, the ML model is trained on measured *in vivo* data. However, the training using measured *in vivo* data usually has two limitations: 1) poor-quality ground truth data; and 2) insufficient training data. Previous attempts have been made to generate ground truth data using the multiple-pool Lorentzian fitted APT effect from each voxel, but the low SNR of CEST signals from a single voxel makes it unsuitable for reliable ground truth data for training. In addition, although models trained on a limited number of patients or healthy subjects can be generalized to unseen patients, improving generalizability requires more training data. First, APT effects vary across different disease types and stages, emphasizing the need for a broader range of training data (8-11). Second, variations in other sample parameters such as water relaxations, amine CEST, nuclear Overhauser enhancement (NOE), and MT effect add complexity to the prediction process in different diseases (28-31). Training data that encompass these diverse features would facilitate more robust predictions. However, acquiring such data is impractical.

The use of synthetic data generated through simulations has become a valuable approach for addressing challenges related to data collection, data annotation (ground truth), and data quality assurance (32,33). In the field of MRI, synthetic data have been successfully employed to train ML models for quantitative parameter prediction (34-43). Synthetic data is categorized as fully synthetic or partially synthetic data. Fully synthetic data solely consists of simulated data, while partially synthetic data involve replacing selected features with simulated data while preserving real data components. In an ideal scenario, fully synthetic data would be generated by numerically simulating the Bloch-McConnell equations, considering all exchanging pools and variations in sample parameters, to closely replicate CEST signals from tissues. However, this proves challenging in practice, due to the unknown values of numerous sample parameters in tissues. Consequently, a wide range of sample parameters must be simulated to ensure the synthetic data encompasses the real sample parameters. This approach may generate training data that contains irrelevant features, potentially leading to overfitting. Furthermore, as the size of the training dataset increases, it may become difficult to fit all the training data into memory. Partially synthetic data offers a more practical solution by combining simulated data with real data components. By incorporating simulated features into real data, partially synthetic data provides a more comprehensive representation of the underlying characteristics of the imaging data. This approach allows for more accurate modeling and prediction of quantitative parameters using ML models.

In this study, we introduced a novel platform that integrates simulated components and measured components to generate partially synthetic CEST data. This data encompasses a wide range of pathological changes and provides accurate ground truth information, addressing the limitations of ML CEST imaging using either the measured data or fully synthetic data. First, we validated the capability of this platform to generate partially synthetic CEST data that accurately mimics real data. Second, we evaluated the accuracy of predicting the APT effect using a ML model trained on the partially synthetic CEST data. Finally, we compared the performance of the ML model trained on the partially synthetic CEST data with that trained directly on measured in vivo data and fully synthetic CEST data, to demonstrate the advantages of our method.

## METHODS

### Animal preparation

Eight rats with 9L tumors were included in this study. For the induction of brain tumor, each rat was injected with $1 \times 10^5$ 9L glioblastoma cells and imaged after 15-20 days. All rats were immobilized and anesthetized with a 2%/98% isoflurane/oxygen mixture during data acquisition. Respiration was monitored to be stable, and a constant rectal temperature of 37°C was maintained throughout the experiments using a warm-air feedback system (SA Instruments, Stony Brook, NY, USA). All animal procedures were approved by the Animal Care and Usage Committee of Vanderbilt University Medical Center.

### MRI

All measurements were performed on a Varian DirectDrive$^{TM}$ horizontal 9.4 T magnet with a 38-mm Litz RF coil (Doty Scientific Inc. Columbia, SC). CEST measurements were performed by applying a continuous-wave CEST sequence with a 5s-rectangular saturation pulse followed by single-shot spin-echo echo planar imaging (SE-EPI) acquisition with repetition time (TR) of 7s. The time of echo (TE) was 27ms. Z-spectra were acquired with RF frequency offsets ($\Delta\omega$) at ±4000, ±3500, ±3000, and from -2000 to 2000 Hz, with a step of 50 Hz (-10 to 10 ppm on 9.4T) and an RF saturation power ($\omega_1$) of 1μT. Control image intensity ($S_0$) was obtained by setting $\Delta\omega$ to 100 kHz (250 ppm on 9.4T). The acquisition of a Z-spectrum took around 15mins. MT pool size ratio ($f_m$) and the observed water longitudinal relaxation time ($T_{1obs} = 1/R_{1obs}$) were measured using a selective inversion recovery (SIR) method (44). All images were acquired with a matrix size of 64 × 64, a field of view of 30 × 30mm$^2$, and one acquisition.

### Quantification metrics

To isolate the target CEST/NOE effect from confounding factors, a reference signal ($S_{ref}$) that contains contributions only from the confounding factors, excluding the target effect, and a label signal ($S_{lab}$) that contains all contributions are typically obtained. Conventionally, a chemical exchange saturation transfer ratio (CESTR) metric is calculated by subtracting $S_{lab}$ from $S_{ref}$. The CESTR is given by,

$$CESTR(\Delta\omega) = \frac{S_{ref}(\Delta\omega)}{S_0} - \frac{S_{lab}(\Delta\omega)}{S_0} \quad (1)$$

Additionally, an apparent exchange-dependent relaxation (AREX) metric (45) which inversely subtracts $S_{lab}$ from $S_{ref}$ with $T_{1obs}$ normalization is given by,

$$AREX(\Delta\omega)=\left(\frac{S_0}{S_{lab}(\Delta\omega)} - \frac{S_0}{S_{ref}(\Delta\omega)}\right)R_{1obs}(1+f_m) = R_{ex}(\Delta\omega) \quad (2)$$

where $R_{ex}$ represents the exchanging/coupling effect in the rotating frame. For slow exchanging/coupling pools (e.g. APT and NOE), $R_{ex}$ can be described by (46)

$$R_{ex}(\Delta\omega) = \frac{f_s k_{sw} \omega_1^2}{\omega_1^2 + (R_{2s}+k_{sw})k_{sw} + \frac{(\Delta\omega-\Delta)^2 k_{sw}}{R_{2s}+k_{sw}}} \quad (3)$$

where $f_s$ and $k_{sw}$ are the solute concentration and solute-water exchange rate, respectively; $\Delta$ is the solute resonance frequency; and $R_{2s}$ is the solute transverse relaxation rate.

**Multiple-pool Lorentzian fit**

The multiple-pool Lorentzian fit approach was used to process the CEST Z-spectra to isolate each pool. The model function for each Lorentzian is defined as:

$$\frac{S(\Delta\omega)}{S_0} = 1 - \sum_{i=1}^{N} L_i(\Delta\omega) \quad (4)$$

Where, S represents CEST image intensity. $L_i(\Delta\omega)=A_i/(1+(\Delta\omega-\Delta_i)^2/(0.5W_i)^2)$, represents a Lorentzian line with a central frequency offset from water ($\Delta_i$), full width at half maximum ($W_i$), and peak amplitude ($A_i$). N is the number of fitted pools. In the brain, several pools have been reported, including amide at 3.5ppm (6), amine at 3ppm (47,48), guanidinium at 2ppm (49,50), hydroxyls below 1ppm (51), water, NOE at -1.6ppm (NOE(-1.6)) (52,53), NOE at -3.5ppm (NOE(-3.5)) (31,54), and MT at -2.3ppm (55). However, due to their proximity in frequency offsets and broad peak shapes, it is challenging to separate amine, guanidine, and hydroxyls from each other. Therefore, for processing the CEST Z-spectra, they are typically treated as a single amine pool centered at 2ppm (56). Here, a six-pool Lorentzian fit model including amide ($L_1$), amine ($L_2$), water ($L_3$), NOE(-1.6) ($L_4$), NOE(-3.5) ($L_5$), and MT ($L_6$) was used to process the entire CEST Z-spectra. The fitting was performed to achieve the lowest root mean square (RMS) of residuals between the measured data and model with the least-squares optimization. Supporting Information Table S1 lists the starting points and boundaries of the fit. In the multiple-pool Lorentzian fit model, $S_{lab}/S_0$ was obtained by doing the subtraction between 1 and the sum of all the Lorentzian fits, and $S_{ref}/S_0$ for each exchanging/coupling pool was obtained by doting the subtraction between 1 and the sum of all Lorentzian fits except that of the corresponding pool (57). The multiple-pool Lorentzian fitted APT spectrum was quantified by Eq. (1).

**Generation of partially synthetic CEST data**

In a multiple-pool model including CEST/NOE, DS, and MT, the CEST signal has been descripted by the inverse summation of these effects in the rotating frame (45,46). This relationship enables us to effectively incorporate the desired combination of simulated and measured components to create the partially synthetic CEST data. In this study, we made modifications to this relationship and formulated Eq. (5) to generate the partially synthetic CEST signals.

$$\frac{S(\Delta\omega)}{S_0} = \frac{R_{1obs}\cos^2\theta}{R_{eff}(\Delta\omega) + \frac{R_{ex}^{APT}(\Delta\omega)}{1+r_{MT}fm} + \frac{R_{ex}^{NOE}(\Delta\omega)}{1+r_{MT}fm} + r_{amines}\frac{R_{ex}^{amines}(\Delta\omega)}{1+r_{MT}fm} + r_{MT}R_{ex}^{MT}(\Delta\omega)} \quad (5)$$

where $R_{eff}$ is the effective water relaxation in the rotating frame; $R_{ex}^{APT}$, $R_{ex}^{NOE}$, $R_{ex}^{amines}$, and $R_{ex}^{MT}$ are APT, NOE, amine CEST, and MT effects in the rotating frame, respectively (45,46); and $r_{\text{amines}}$ and $r_{\text{MT}}$ are two scaling factors. Supporting information Table S2 describes each parameters.

Specifically, we used $R_{ex}^{amines}$ and $R_{ex}^{MT}$ components from measurements, as well as $R_{ex}^{APT}$, $R_{ex}^{NOE}$, and $R_{eff}$ components from simulations, to generate partially synthetic CEST data. The $R_{ex}^{amines}$ component was quantified by the AREX metric in Eq. (2), employing the multiple-pool Lorentzian fitted $S_{ref}$ for amines. The $R_{ex}^{MT}$ component was calculated using the formula $R_{1obs}L_6/(1-L_6)$, where $L_6$ is also obtained from the multiple-pool Lorentzian fit (see Supplementary Information Theory for the derivation). The $R_{ex}^{APT}$ and $R_{ex}^{NOE}$ components were calculated using Eq. (3). The $R_{eff}$ component was obtained using the following equations:

$$R_{eff}(\Delta\omega) = R_{1obs}\cos^2\theta + R_{2w}\sin^2\theta \quad (6)$$

$$\text{Cos}^2\theta = \frac{\Delta\omega^2}{\omega_1^2+\Delta\omega^2}; \qquad \sin^2\theta = \frac{\omega_1^2}{\omega_1^2+\Delta\omega^2}$$

where $R_{1obs}$ was calculated using

$$R_{1obs} = \frac{R_{1w}+r_{MT}f_mR_{1M}}{1+r_{MT}f_m} \quad (7)$$

in which $R_{1M}$ is the semisolid longitudinal relaxation rate. $f_m$ is obtained from measurement to align with $R_{ex}^{MT}$.

Sample parameters including $f_s$, $k_{sw}$, $T_1$, and $T_2$, along with the scaling factors r for each pool were varied to generate diverse training data, as detailed in supporting information Table S3. The target CESTR quantified APT spectrum was obtained by the subtraction of two partially synthetic CEST Z-spectra with/without the amide pool ($R_{ex}^{APT}$) using Eq. (5).

**Generation of tissue-mimicking CEST data**

To evaluate the accuracy of the ML model trained on the partially synthetic CEST data, we generated tissue-mimicking CEST data that provided ground truth information. The tissue-mimicking CEST Z-spectra were created through numerical simulation of the Bloch-McConnell equation, using the same sequence parameters as those employed in the MRI experiments. To simulate the major tissue components found in the brain, we employed a seven-pool model that included amide at 3.5ppm, amine at 3ppm, guanidinium at 2ppm, water, NOE(-1.6), NOE(-3.5), and MT (56). For the purpose of testing other methods, we varied sample parameters including $f_s$, $k_{sw}$, $T_1$, and $T_2$ for each pool, as detailed in supporting information Table S4. The values of $f_s$, $k_{sw}$, and $T_2$ for water and amide pool, as well as the value of $f_s$ for NOE(-3.5), were different from those used to create the partially synthetic data. Additionally, the values of $f_s$, $k_{sw}$, and/or $T_2$ for guanidine, amine, and MT were varied to introduce variations in the amplitude and width of their CEST/MT effects. The ground truth CESTR-quantified APT spectrum for this tissue-mimicking data was obtained by the subtraction of two CEST Z-spectra with/without the amide pool through the simulation of the Bloch equations.

**Generation of fully synthetic CEST data**

The fully synthetic CEST Z-spectra were also created through numerical simulation of the Bloch-McConnell equation, using the same sequence parameters as those employed in creating the partially synthetic data and followed the same models employed for the tissue-mimicking data. The target CESTR-quantified APT spectrum (also ground truth) for this fully synthetic data was obtained by subtracting two CEST Z-spectra with/without the amide pool through the simulation of the Bloch equations.

**Data analysis**

**1)** To validate the feasibility to use Eq. (5) for generating accurate partially synthetic CEST signals, we used the average of each multiple-pool Lorentzian fitted CEST/MT component, $R_{1obs}$, and $f_m$ ($r_{amines}$=0; $r_{MT}$=0) from the measurements on the eight rat brains to generate a synthetic CEST Z-spectrum and compared this spectrum with the original measured CEST Z-spectrum. It is worth noting that, in this validation process, the $R_{ex}^{APT}$ and $R_{ex}^{NOE}$ components were quantified by the AREX metric, employing the multiple-pool Lorentzian fitted $S_{ref}$ for them, instead of using simulations. Additionally, the $R_{eff}$ component was calculated using the formula $R_{1obs}S_{ref_w}/S_{lab}$, where $S_{ref_w}$ represents the reference signal from the water pool, but not the simulation (see Supplementary Information Theory for the derivation).

**2)** To validate the accuracy of the ML method using the partially synthetic CEST data to predict APT signals, we applied it to process the tissue-mimicking data and compared the results with the conventional multiple-pool Lorentzian fitted results. First, from a few randomly selected Z-spectra within the tissue-mimicking data, we obtained the multiple-pool Lorentzian fitted components $R_{ex}^{amines}$ and $R_{ex}^{MT}$. These fitted components, along with simulated $R_{eff}$, $R_{ex}^{APT}$ and $R_{ex}^{NOE}$, were used to generate the partially synthetic data. This data is called partially synthetic data generated with measured components from tissue-mimicking data to distinguish it from the partially synthetic data generated with measured components from *in vivo* data. The target CESTR quantified APT spectrum was obtained by the subtraction of two partially synthetic CEST Z-spectra with/without the amide pool ($R_{ex}^{APT}$) using Eq. (5). Next, we trained the ML model using this partially synthetic data. The trained ML model was then applied to predict the CESTR quantified APT spectra for 1000 testing samples of tissue-mimicking data. We obtained the average difference between the ML-predicted APT spectra and the ground truth APT spectra of the tissue-mimicking data within the frequency range of 2ppm to 5ppm. This average difference (i.e., loss) served as a measure of the ML model's accuracy. Similarly, we obtained the average difference between the multiple-pool Lorentzian fitted APT spectra and the ground truth APT spectra from the tissue-mimicking data. By comparing these average differences from these two methods, we were able to assess whether the ML approach using partially synthetic data outperforms the multiple-pool Lorentzian fit.

**3)** To evaluate the advantage of the ML method using the partially synthetic CEST data compared to using the *in vivo* data alone, we conducted two types of comparisons. In type 1, from an average of randomly selected Z-spectra from a varying number of voxels (i.e, 50, 100, 500 voxels) and of all voxels (2157 voxels) within the measured data from five rats, we obtained the multiple-pool Lorentzian fitted components $R_{ex}^{amines}$ and $R_{ex}^{MT}$. Using these components, we generated partially synthetic data and trained the ML model on these data. We then applied the model to predict the APT spectrum on the other three rats. Additionally, we directly trained the ML model on *in vivo* Z-spectra from the corresponding voxels to those used in creating the partially synthetic data, using the multiple-pool Lorentzian fitted APT spectrum for each voxel as the target. We also augmented the training data created from all voxels of the five rat brains (2157 voxels) by averaging each pair of Z-spectra, resulting in an expansion of the dataset to nearly 2000 times its original size. The multiple-pool Lorentzian fitted APT spectra from the augmented Z spectra were used as targets. This trained ML model was employed to predict APT spectra from the other three rats. This type 1 comparison can also demonstrate the model's generalizability to other subjects. In type 2, from the averaged Z-spectra from all normal tissues or all tumor tissues of the eight rats, we obtained the multiple-pool Lorentzian fitted components $R_{ex}^{amines}$ and $R_{ex}^{MT}$ and generate the partially synthetic data. Additionally, we trained the ML model on all measured Z-spectra from all voxels in either tumors or normal tissues, using the multiple-pool Lorentzian fitted APT spectrum for each voxel as the target. Furthermore, we augmented the training data created from all voxels in normal tissues (1923 voxels) or in tumor tissues (491 voxels) of the eight rats by averaging each pair of Z-spectra, resulting in an expansion of the dataset to nearly 2000 or 500 times its original size respectively. The target was created by using the multiple pool Lorentzian fitted APT spectrum from each of the augmented Z-spectrum. We then applied our NN model to process the data from other parts of the rat brains. This type 2 comparison can also show the model's generalizability to other pathologies.

**4)** To evaluate the advantage of the ML method using the partially synthetic CEST data compared to using fully synthetic data, we also conducted two types of comparisons. In both these two types, the multiple-pool Lorentzian fitted components $R_{ex}^{amines}$ and $R_{ex}^{MT}$ were obtained from the average of 50 randomly selected measured Z-spectra, which were then used to create partial synthetic data. In type 1, the fully synthetic data were generated with the same sample size as the partially synthetic data for fair comparison. Specifically, $f_s$, $k_{sw}$, and $T_2$ for water, amide, and NOE(-3.5) pools were set the same as those used in creating the partial synthetic data. The $f_s$ of amine and MT were set to be $0.003 \cdot r_{amine}$ and $0.1 \cdot r_{MT}$, respectively. In type 2, the fully synthetic data were generated to have three times the size of the partially synthetic data by adding more simulations with varied $f_s$ of guanidine, as detailed in Supporting information Table S5.

**Machine Learning (ML)**

Fig. 1 shows a flowchart of the machine learning workflow, along with the utilized neural network (NN). The input data consisted of the Z-spectra with $\Delta\omega$ ranging from -10ppm to -5ppm, -0.5ppm to 0.5ppm, and 2.5ppm to 10ppm at 9.4T (39 data points). The data from -5ppm to -0.5ppm and 0.5ppm to 2.5ppm were excluded since their impact on the CEST signal at 3.5ppm is negligible. The output data consisted of the amplitude (*A*) and width (*W*) of the target amide CESTR peak. *A* was calculated by setting $\Delta\omega$ to $\Delta$ in the amide CESTR peak, and *W* was obtained as the full width at half maximum (FWHM) of the amide CESTR peak. The ML-predicted APT spectrum was generated by applying a Lorentzian function based on the predicted *A* and *W* (25). An NN model with two dense layers, each comprising 100 neurons and an output layer with two nodes was employed to predict the *A* and *W* values. The rectified linear unit activation (ReLU) activation function applied after each dense layer and mean square error (MSE) loss function were utilized. The NN model was trained on four types of data - partially synthetic data with measured components from the tissue-mimicking data, partially synthetic data with measured components from *in vivo* data, *in vivo* data, and the fully synthetic data. For the training using partially synthetic data and the fully synthetic data, gaussian noises with a standard deviation of 0.01 were added to the Z-spectra to simulate real tissue signals. The NN model was implemented in MATLAB R2022a using the Adam optimizer for 4000 epochs, with a learning rate of $1\text{x}10^{-3}$ and a batch size of 64. Supporting Information Fig. S2, shows the training and validation loss versus the number of epochs. Training was stopped when validation loss began to increase, while training loss continued to decrease. The training took ~3h while using partially synthetic and fully synthetic data and ~20 mins for *in vivo* data without augmentation. The ML prediction of the APT map for a rat brain took approximately 1.6 seconds on a Dell System with Intel(R) Xeon(R) CPU E5-1607 v2 3.00 GHz processor.

**Statistics**

The tumor ROIs were outlined based on the $f_m$ map, with values less than a threshold of 7%. The contralateral normal tissue ROIs were chosen to mirror the tumor ROIs. Student's t-tests were employed to compare the ROI-averaged signals. We considered $P < 0.05$ to be statistically significant. The normal tissue ROIs, used for selecting training data, were chosen by subtracting the whole brain ROI from the tumor ROI.

## RESULTS

**Validation of the feasibility to generate partially synthetic CEST data using Eq. (5)**

Figs. 2a and 2b show the multiple-pool Lorentzian fitted $R_{eff}$, $R_{ex}^{APT}$, $R_{ex}^{NOE}$, $R_{ex}^{amines}$, and $R_{ex}^{MT}$ spectra from the averaged Z-spectrum of all normal tissues. Fig. 2c shows a comparison of the synthetic Z-spectrum generated using Eq. (5) with the original measured Z-spectrum. The close resemblance between the synthetic and measured Z-spectra indicates the feasibility of constructing CEST signals using different multiple-pool Lorentzian fitted components according to Eq. (5). Fig. 2d presents a comparison between the measured Z-spectrum and the synthetic Z-spectrum without the APT effect. The discrepancy between the two spectra represents the APT contribution. Thus, by substituting the multiple-pool Lorentzian fitted APT components with simulated APT components, partially synthetic CEST data can be generated. The huge peak at 0ppm in Fig. 2b is due to the inverse of a very small value at 0ppm. Supporting information Fig. S1 shows these spectra from all tumors, showing close resemblance between the synthetic and measured Z-spectra as well.

**Validation of the accuracy of the ML method using partially synthetic CEST data generated with measured components from the tissue-mimicking data.**

Fig. 3a displays a representative Z-spectrum from the tissue-mimicking data. Fig. 3b presents a comparison of the corresponding APT spectra from the ML prediction, multiple-pool Lorentzian fit, and the ground truth. This ML model was trained on partially synthetic data, with the measured components fitted from one randomly selected Z-spectrum within tissue-mimicking data. It can be observed that the ML predicted APT spectrum closely resembles the ground truth, whereas the Lorentzian fitted APT spectrum exhibits a difference. Fig. 3c illustrates the comparison of losses between the ML method and the multiple-pool Lorentzian fit for all testing samples. The results indicate that the ML method accurately predicts the APT spectra of all testing samples, with a mean loss less than $8.9 \times 10^{-4}$. In contrast, the multiple-pool Lorentzian fit shows significant bias in some of the testing samples. Supporting information Fig. S3 compares losses between the ML method and the multiple-pool Lorentzian fit, with the measured components used to generate the partially synthetic data, fitted from ten other randomly selected Z-spectra, as well as from the average of 50, 100, 500, and 1000 randomly selected Z-spectra, within the tissue-mimicking data. The results consistently demonstrate the accuracy of the ML method using partially synthetic data in predicting APT, while the multiple-pool Lorentzian fit exhibits overestimation in some of the testing samples.

**Comparison of the ML model trained on partially synthetic CEST data generated with measured components from in vivo data, directly on the in vivo data, and on fully synthetic CEST data.**

Fig. 4 displays the average CEST Z-spectra from the three testing rats, and the corresponding average APT spectra from the multiple-pool Lorentzian fit, ML prediction using the partially synthetic data and the *in vivo* data with the type 1 selection as well as using the *in vivo* data with data augmentation. Fig. 5 shows the average CEST Z-spectra from the eight rats, and the corresponding averaged APT spectra from the multiple-pool Lorentzian fit, ML prediction using the partially synthetic data and the in vivo data with the type 2 selection, as well as ML prediction using fully synthetic data with the type 1 and type 2 simulations. Supporting information Fig. S4 and S5 show these plots in Fig. 4 and 5, respectively, with the standard deviation across subjects. The ML predicted APT spectra, trained using all types of partially synthetic data, are roughly consistent, suggesting that our method does not depend on the size of samples for extracting the measured components. In addition, the amplitude is lower than the multiple-pool Lorentzian fitted APT spectra, due to the overestimation of APT using the multiple-pool Lorentzian fit as shown in Fig. 3. In contrast, the ML-predicted APT spectra trained using all types of in vivo data is not consistent. Generally, for training the ML using in vivo data, the higher the voxel count, the closer is predicted APT spectra to the multiple-pool Lorentzian fit, indicating that this type of training requires a larger dataset. The ML-predicted APT spectra trained using all types of fully synthetic data also shows significant deviations from the multiple-pool Lorentzian fitted APT spectra. Specifically, type 1 simulations yield much lower ML-predicted APT spectra, while type 2 simulations yield much higher ML-predicted APT spectra compared to the multiple-pool Lorentzian fitted APT spectra. These findings suggest that training with fully synthetic data heavily relies on the simulation models employed. Supporting information Fig. S6 and Fig. S7 display the pixel-wise regression plots between the multiple-pool Lorentzian fitted APT amplitude and the predicted APT amplitude from all the ML models presented in Fig. 4 and Fig. 5, respectively. Notably, while the $R^2$ values for the training using the partially synthetic data are very high (>0.98), indicating successful training, the $R^2$ values for the training using *in vivo* data and fully synthetic data are considerably lower, indicating unsuccessful training. Additionally, the $R^2$ values gradually increase (0.596, 0.617, 0.633, 0.701, and 0.859) for ML using in vivo data with type 1 selection of 50, 100, 500, 2157 voxels and with data augmentation, as well as increase from 0.776 and 0.831 to 0.903 and 0.958 for ML with type 2 selection of tumors and normal tissues after using data augmentation, confirming that it rely on significantly larger training data to achieve higher predictive accuracy.

Fig. 6 shows the APT amplitude maps from the three testing rat brains using the Lorentzian fitting and ML prediction. For ML prediction, partially synthetic data and in vivo data with type 1 selection are used, along with in vivo data with data augmentation. Fig. 7 shows the APT amplitude maps from the one rat brain using the Lorentzian fitting and ML prediction using partially synthetic data, *in vivo* data with the type 2 selection, *in vivo* data with the type 2 selection with data augmentation, as well as ML prediction using fully synthetic data with the type 1 and type 2 simulations. Supporting information Fig. S8 shows the APT amplitude maps from the rest seven rat brains. Supporting information Fig. S9 shows the anatomy images from the eight rat brains. The ML-predicted APT map, trained using all types of partially synthetic data, demonstrates good image quality. However, ML prediction using *in vivo* data and fully synthetic data does not yield satisfactory image quality in terms of the contrast between tumor and normal tissues when compared to the anatomy images. Notably, the ML-predicted APT map trained using *in vivo* data with data augmentation exhibits good image quality but with reduced contrast between tumor and normal tissues. This suggests that even with augmentation, some features may not be adequately captured in the training data.

**Evaluation of the ML predicted APT imaging to detect tumors**

In Fig. 8, the statistical difference in the $R_{1obs}$, $f_m$, the ML predicted APT amplitude using partially synthetic data with measured components from the type 1 selection of 50 voxels, and multiple-pool Lorentzian fitted APT amplitude are presented. All these parameters exhibit significant differences between tumor and contralateral normal tissues. Tumors display significantly higher

APT amplitudes compared to contralateral normal tissues, which is consistent with previous findings (6,7). Notably, the multiple-pool Lorentzian fitted APT amplitudes from both tumor and contralateral normal tissues are higher than those obtained from the ML prediction using the partially synthetic data, which is due to the overestimation of APT effect using the fitting approach as shown in Fig.3.

## DISCUSSION

We have introduced a novel platform for generating partially synthetic CEST data by combining simulated components with measured components. Utilizing this approach in machine learning offers several advantages: 1) The ability to create training data with any sample parameters (e.g., APT/NOE $f_s$, $k_{sw}$ and $R_{2s}$, the amplitude of amine CEST and MT effects) by adjusting the parameters of the simulated components and scaling factors of the measured components. This addresses the constraint of limited training data when using only measured data in machine learning. 2) By incorporating the measured components, our approach maintains high simulation fidelity. In biological tissues, there are multiple amines from 2-3ppm and hydroxyls below 1ppm with broad ranges of exchange rate (50,51,58-62). Simulating all these pools accurately is challenging due to the unknown parameter ranges. However, the average contributions of these pools can be roughly reflected by the fitted $R_{ex}^{amines}$, as shown in Fig. 2b. Additionally, the broad nature of MT pool makes it difficult to estimate its peak width and offset accurately, but these sample parameters do not need to be simulated when using the fitted $R_{ex}^{MT}$ spectrum. This addresses the challenge of inaccurate simulation models when using fully synthetic data in machine learning. 3) The partially synthetic data provides high-quality ground truth data, ensuring reliable training. In contrast, the training using Z-spectra from each voxel to obtain ground truth data suffers from the low SNR. In addition, the use of the multiple-pool Lorentzian fit to obtain ground truth data may have bias. Supporting information Fig. S10 demonstrates a correlation between the amine $f_s$ and the loss between the multiple-pool Lorentzian fit and the ground truth, indicating the influence of the amine CEST effect on the accuracy of the multiple-pool Lorentzian fit. This finding aligns with our previous study showing that the multiple-pool Lorentzian fit performs better when there are no amine pools (56).

It should be noted that we utilized the line shape information of the amine and MT pool to create the partially synthetic data, but not their amplitudes. The amplitude information is compensated for by the scaling factor, ensuring that even if the amplitude is not accurately fitted using the multiple-pool model Lorentzian fit, it does not affect the accuracy of the ML model trained on this partially synthetic data. We did not vary the width of the $R_{ex}^{amines}$ and $R_{ex}^{MT}$ peaks when creating the training data in this paper. However, it is theoretically possible to adjust the widths by varying their Lorentzian fitted widths, which would enable the creation of more training data covering a broader range of variations in different pathologies. Although we did not explore this approach in our study, we conducted validation using tissue-mimicking data that contained varied widths of the amine and MT effects. We found that our method successfully predicted all the testing tissue-mimicking data, indicating its robustness and effectiveness even with varying widths of the amine and MT peaks. The $R_{ex}^{APT}$ in Eq. (5) can also be obtained from the multiple-pool Lorentzian fitted $R_{ex}^{APT}$ by tuning the peak amplitude and width. Here, we used Eq. (3) to calculate $R_{ex}^{APT}$ because it is more directly related to the underlying exchange parameters, so it is easier to determine the range of the simulation parameters by searching the literature.

To evaluate the robustness of the NN trained on the partially synthetic data, we reproduced Fig. 3 with varying levels of noise in supporting information Fig. S11. It can be found that our method shows accurate prediction even with low-SNR data.

As a proof of concept, the new platform to create the synthetic CEST data using Eq. (5) is based on the acquisition of steady-state signals. Theoretically, the method can be extended to create non-steady-state synthetic CEST signal ($S^{nss}$) using (46,63),

$$\frac{S^{nss}(\Delta\omega)}{S_0} = \left(1 - \frac{S(\Delta\omega)}{S_0}\right) \exp\left(-R_{1\rho} t_p\right) + \frac{S(\Delta\omega)}{S_0} \quad (8)$$

where, $t_p$ is the total saturation time, and $R_{1\rho}$ is the summation of all components in their rotating frame (i.e., the denominator in Eq. (5)). The multiple-pool Lorentzian fitted components can be obtained from a baseline steady-state acquisition in one or a few subjects. Then, together with the simulated components and Eq. (8), the platform can be extended to non-steady state conditions. In human brains, different tissues such as white matter, gray matter, and cerebrospinal fluid may exhibit different features of the fitted amine and MT effects in proximity to the amide offset. In this case, inclusion of $R_{ex}^{amines}$ and $R_{ex}^{MT}$ from these tissues into the training data, albeit from a limited number of subjects, can encompass all possible features.

In this paper, we predicted the CESTR quantified APT effect, instead of the AREX quantified APT effect, since the CESTR has been widely used previously although the mechanisms of these two quantification metrics is still not yet fully understood. To predict the AREX quantified APT, similar training methods can be also used.

## CONCLUSION

The utilization of partially synthetic data can effectively address the challenges of limited data availability and the absence of high-quality ground truth data in ML methods using measured data. It also provides enhanced fidelity and robustness compared to using fully synthetic data.

## DATA AVAILABILITY STATEMENT

The code for simulations that support the findings of this study are openly available in *CESTLabZu* at https://github.com/CESTlabZu/MachineLearningbasedAPTImaging. The in vivo data used in this study are available on request from the corresponding author upon request.

**Main Document Figures**

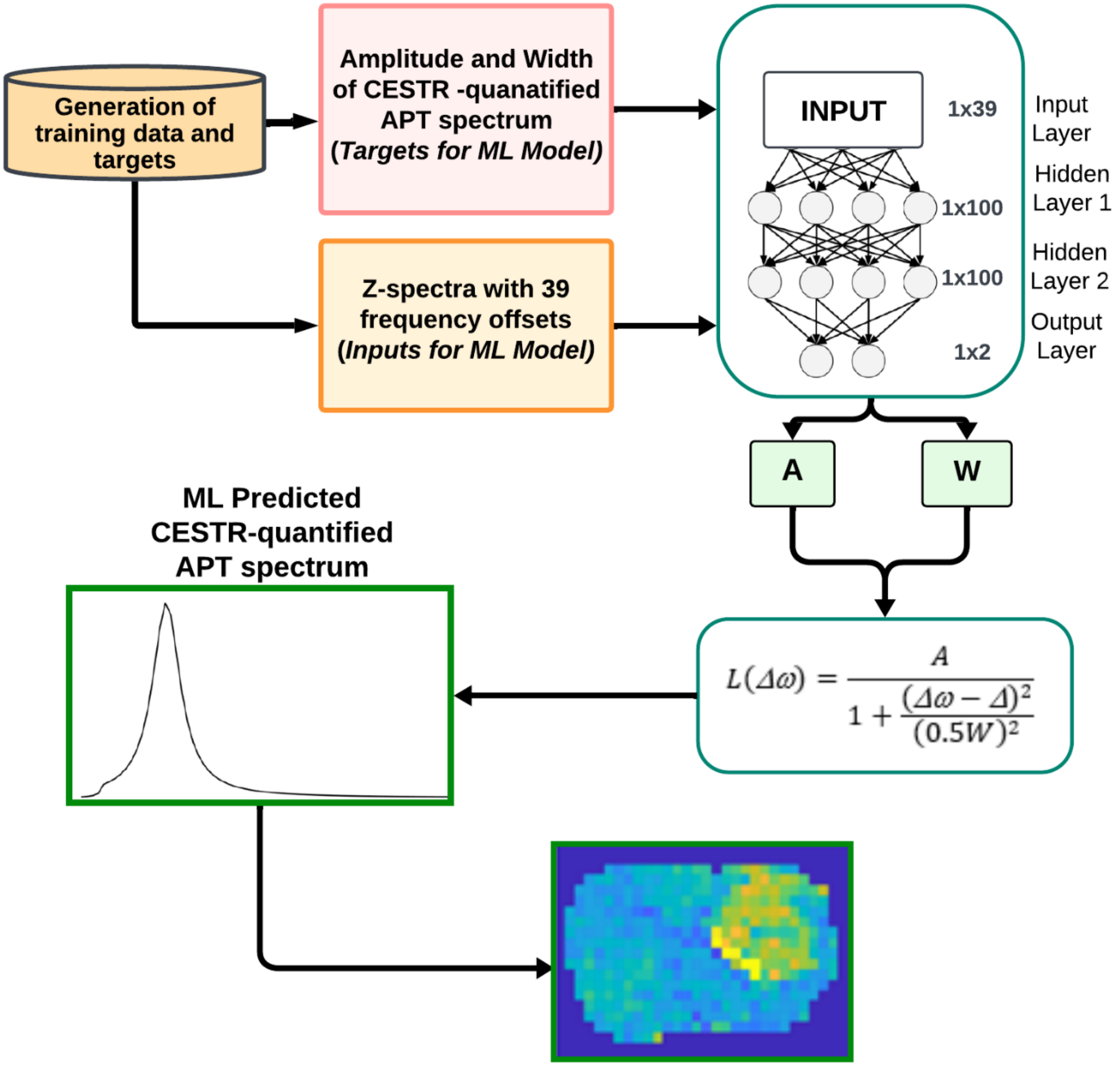

**Fig. 1** Flowchart of the machine learning method.

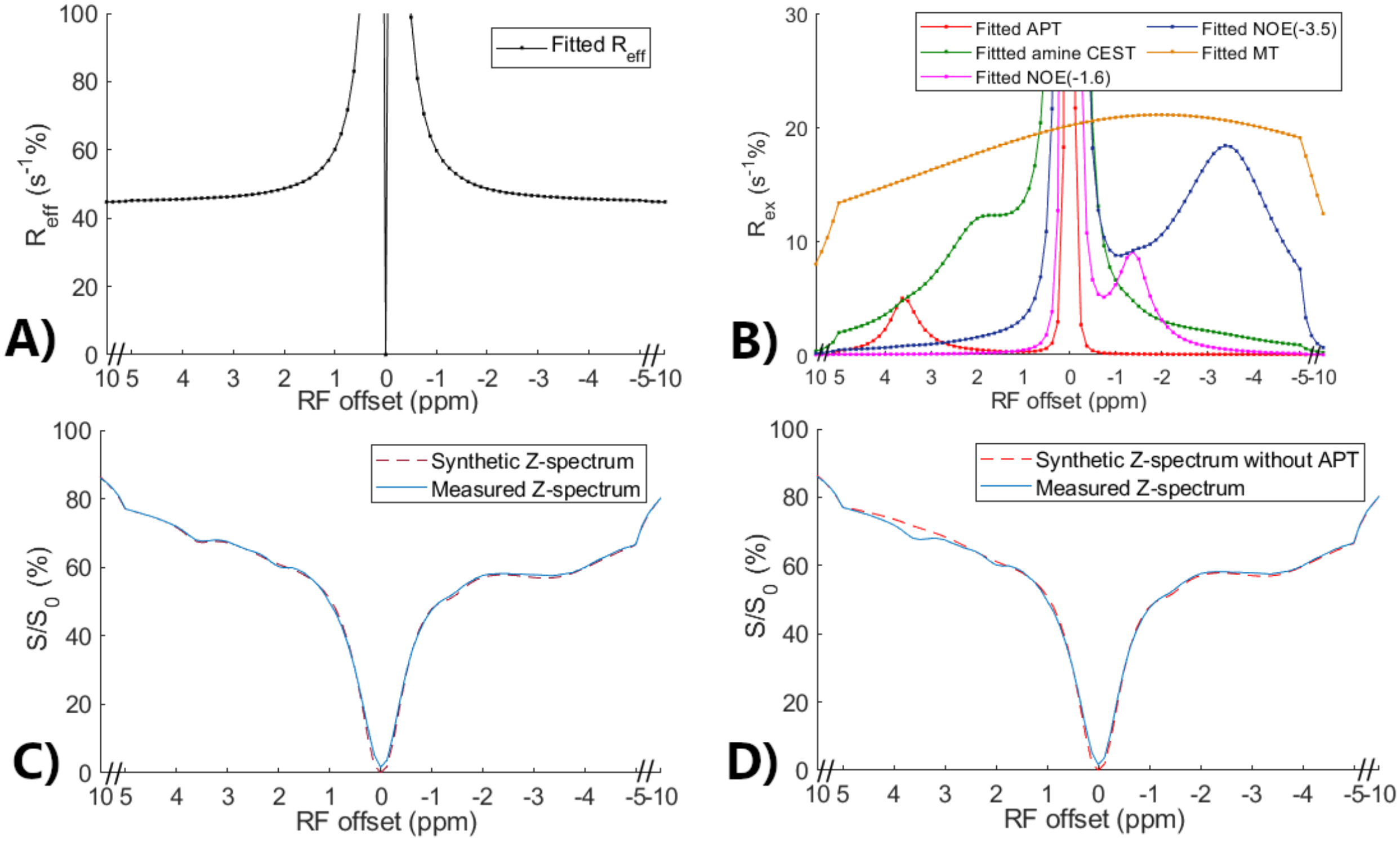

**Fig. 2** Multiple-pool Lorentzian fitted $R_{eff}$ (a) and fitted APT, amine CEST, NOE(-1.6), NOE(-3.5), and MT spectra (b) from the average of the measured CEST Z-spectra in normal tissues in eight rat brains. Comparison between the measured CEST Z-spectra and synthetic Z-spectrum generated using all multiple-pool Lorentzian fitted components (c) as well as between the measured CEST Z-spectra and synthetic CEST Z-spectrum without APT (d).

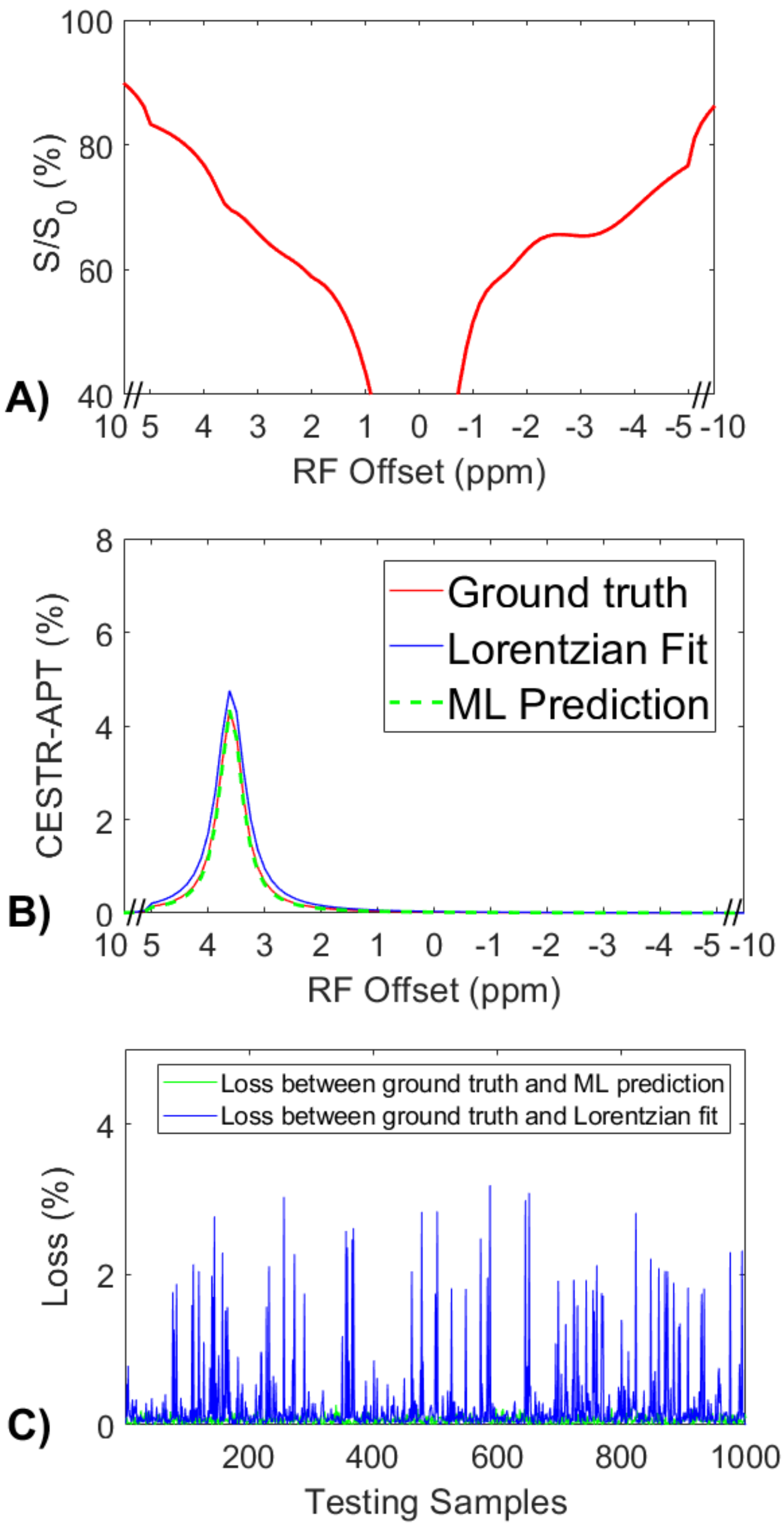

**Fig. 3** (a) A representative Z-spectrum from the tissue-mimicking data. (b) A comparison of the corresponding APT spectra from the ML prediction, multiple-pool Lorentzian fit, and ground truth. (c) A comparison of losses between the ML prediction and the multiple-pool Lorentzian fit for all testing data.

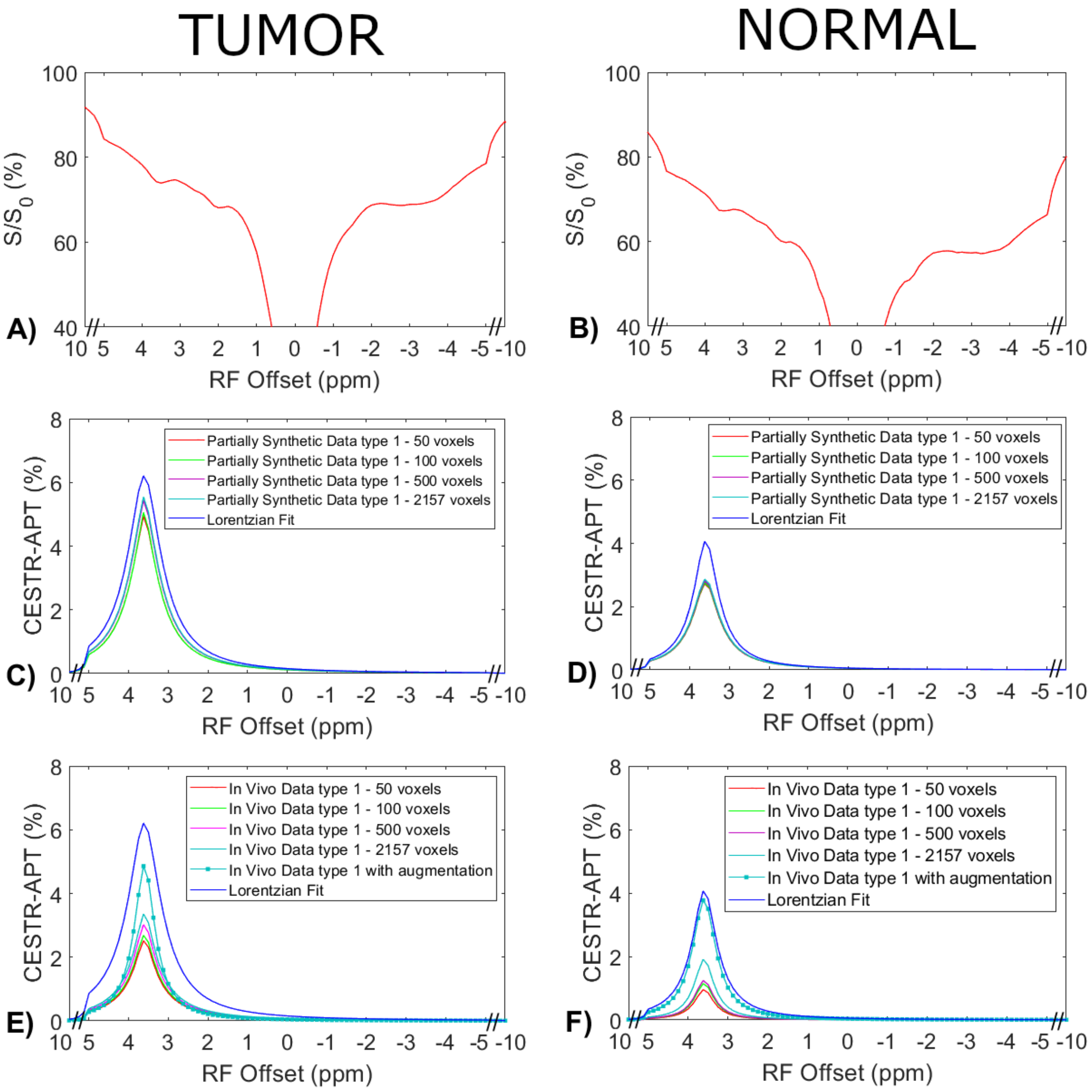

**Fig. 4** Measured CEST Z-spectra from tumors and contralateral normal tissues in the three testing rats (a, b), the corresponding APT spectra from the ML prediction using the partially synthetic data for type 1 selection of 50, 100, 500, and 2157 voxels within five rat brains (c, d), and the corresponding APT spectra from the ML prediction using the measured in vivo data for type 1 selection of 50, 100, 500, and 2157 voxels within five rat brains as well as with data augmentation (e, f). The multiple-pool Lorentzian fitted APT spectra were also plotted in (c-f) for comparison. Data are averaged across different subjects.

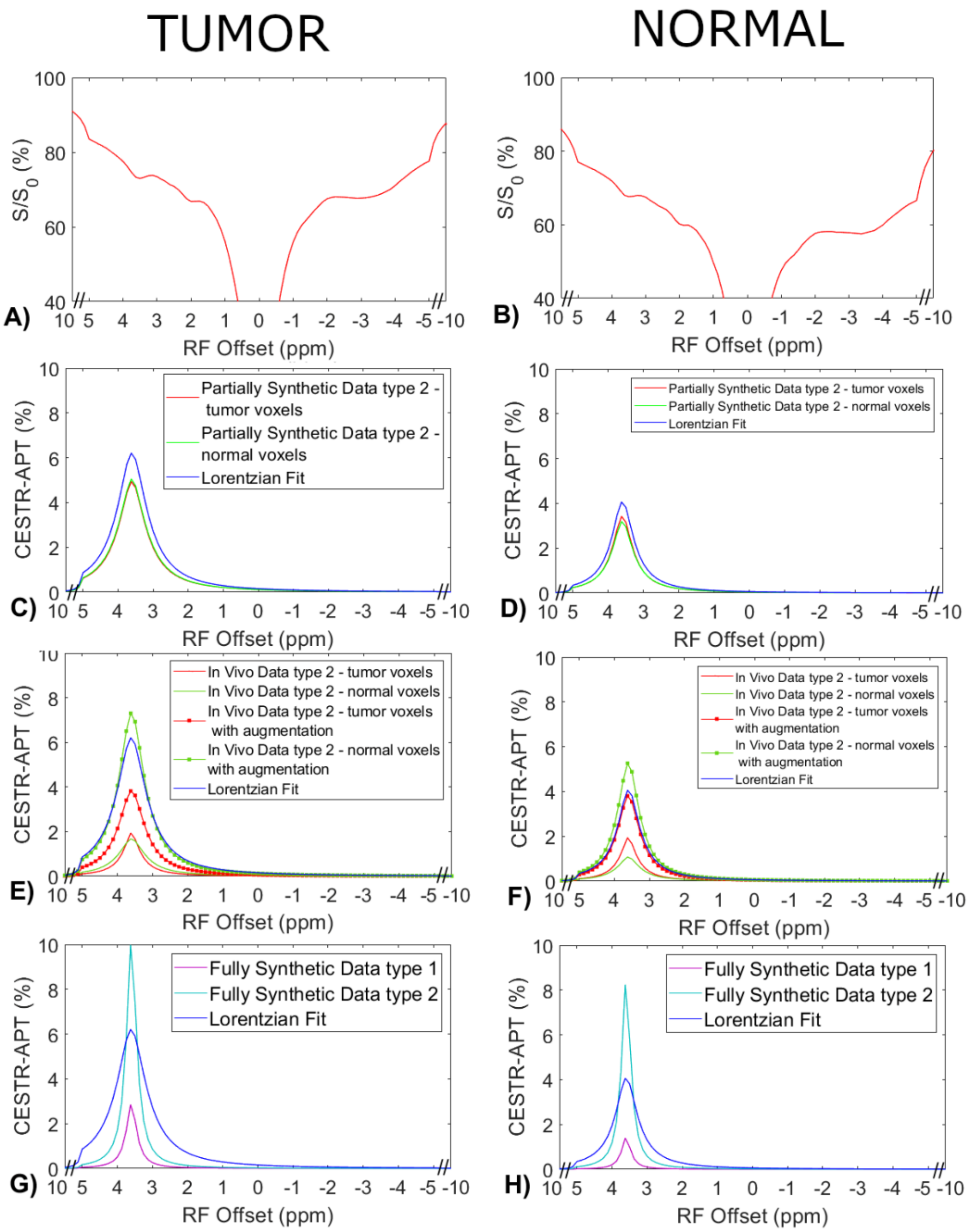

**Fig. 5** Measured CEST Z-spectra from tumors and contralateral normal tissues in the eight rats (a, b), the corresponding APT spectra from the ML prediction using the partially synthetic data for type 2 selection of tumors and normal tissues (c, d), the corresponding APT spectra from the ML prediction using the measured in vivo data for type 2 selection of tumors and normal tissues (e, f), and ML prediction using fully synthetic data with the type 1 and type 2 simulations (g, h). The multiple-pool Lorentzian fitted APT spectra were also plotted in (c-h) for comparison. Data are averaged across different subjects.

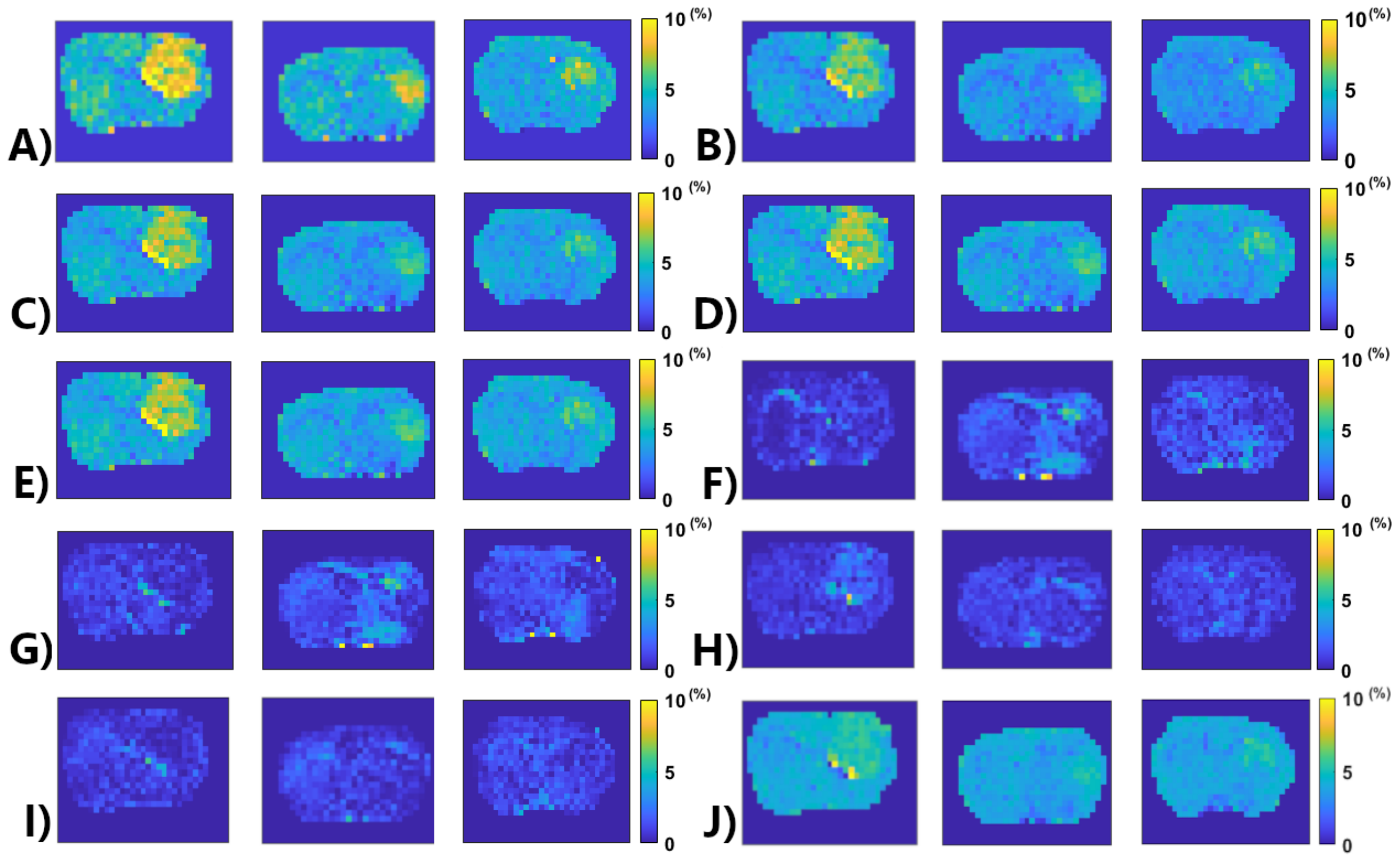

**Fig. 6** APT amplitude maps from three testing rat brains (from left to right – Rat 6, Rat 7, Rat 8) using the Lorentzian fitting (a), ML prediction using partially synthetic data with the measured components from the in vivo data with type 1 selection of 50 (b), 100 (c), 500 (d), and 2157 (e) voxels from five rat brains, ML prediction using the measured in vivo data with type 1 selection of 50 (f), 100 (g), 500 (h), and 2157 (i) voxels from five rat brains as well as ML prediction using the measured in vivo data using all voxels from five rat brains with data augmentation (j).

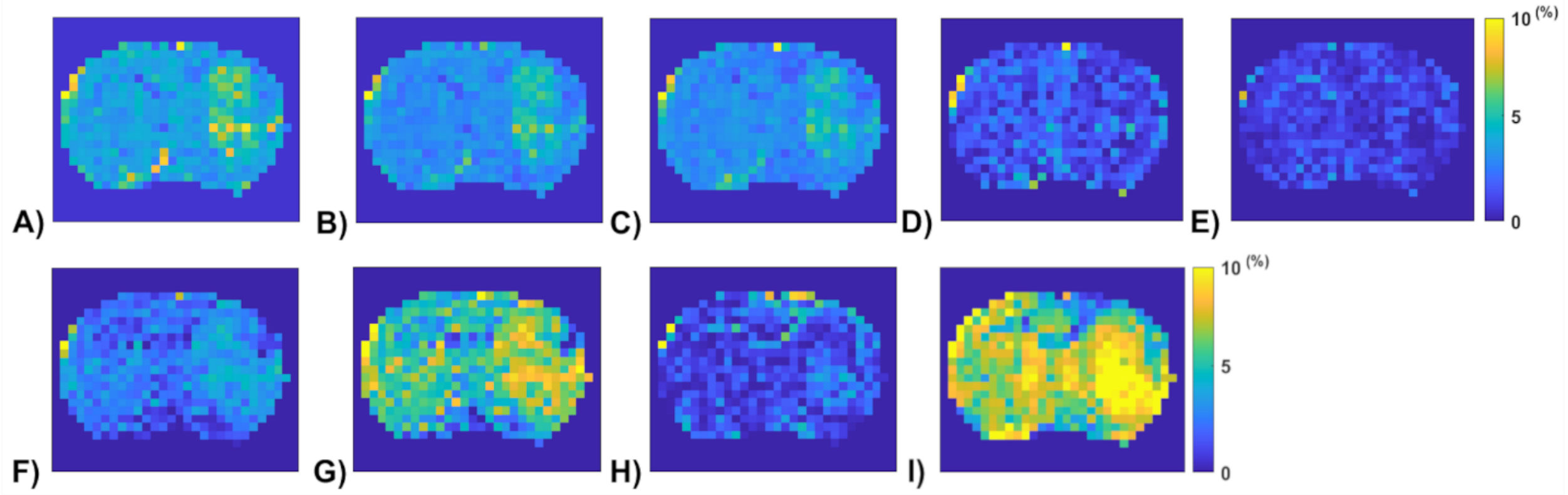

**Fig. 7** APT amplitude maps from the one rat brain (Rat 1) using the Lorentzian fitting (a), ML prediction using partially synthetic data with the measured components from the in vivo data with type 2 selection of tumors (b) and normal tissues (c), ML prediction using the measured in vivo data with type 2 selection of tumors (d) and normal tissues (e), ML prediction using the measured in vivo data with type 2 selection of tumors with data augmentation (f) and normal tissues with data augmentation (g), as well as ML prediction using fully synthetic data with the type 1 (h) and type 2 (i) simulations.

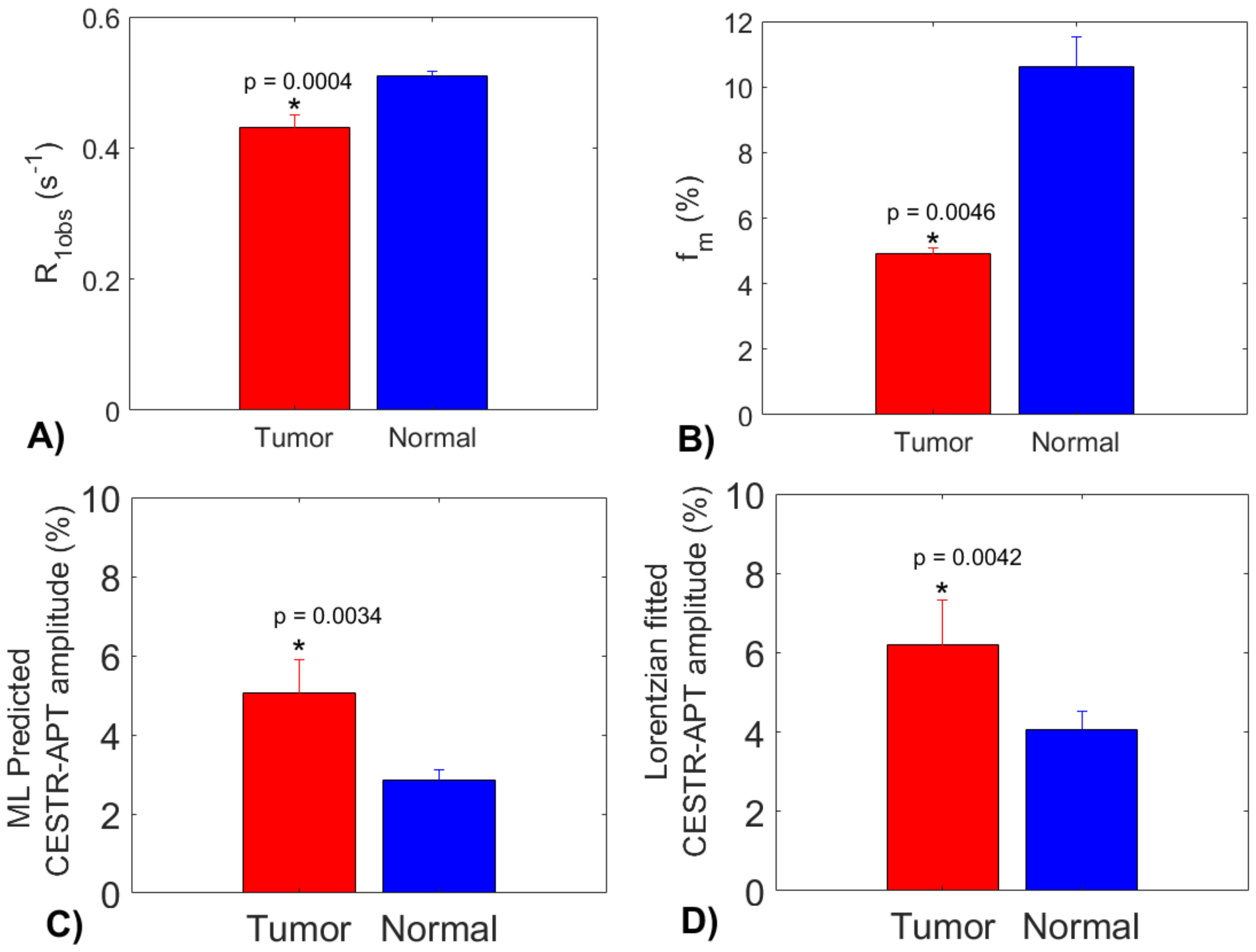

**Fig. 8** Statistical differences in $R_{1obs}$ (a), $f_m$ (b), the ML predicted APT amplitude using partially synthetic data with measured components from the type 1 selection of 50 voxels (c), and the multiple-pool Lorentzian fitted APT amplitude, between tumor and normal tissues in the three testing rat brains.

# Supporting information

**Supporting information Theory:** Multiple-pool Lorentzian fitted $R_{eff}$ and $R_{ex}^{MT}$

In a multiple-pool model including CEST/NOE, DS, and MT effects, these effects acquired in steady state can be described simultaneously by superimposing their rotating frame relaxations, assuming concentrations of the CEST/NOE and MT pools are much less than 1 (1,2). The equation is as follows:

$$R_{1\rho}(\Delta\omega) \approx R_{eff}(\Delta\omega) + \frac{R_{ex}^{CEST}(\Delta\omega)}{1+f_m} + R_{ex}^{MT}(\Delta\omega) \tag{1}$$

$R_{1\rho}(\Delta\omega)$ can be approximated as:

$$R_{1\rho}(\Delta\omega) \approx \frac{S_0 R_{1obs}\cos^2\theta}{S(\Delta\omega)} \tag{2}$$

By substituting $R_{1\rho}$ from Eq. (2) into Eq. (1), we obtain:

$$\frac{S(\Delta\omega)}{S_0} \approx \frac{R_{1obs}\cos^2\theta}{R_{eff}+R_{ex}^{CEST}(\Delta\omega)/(1+f_m)+R_{ex}^{MT}(\Delta\omega)} \tag{3}$$

The signal $S(\Delta\omega)$ in Eq. (3) can also be considered as a label signal ($S_{lab}$). In the multiple-pool Lorentzian fit, the quantity $1-S(\Delta\omega)/S_0$ is first calculated before the fitting. From Eq. (3), this quantity can be further expressed as:

$$1-\frac{S(\Delta\omega)}{S_0} \approx \frac{R_{eff}-R_{1obs}\cos^2\theta+R_{ex}^{CEST}(\Delta\omega)/(1+f_m)+R_{ex}^{MT}(\Delta\omega)}{R_{eff}+R_{ex}^{CEST}(\Delta\omega)/(1+f_m)+R_{ex}^{MT}(\Delta\omega)}$$

$$= \underbrace{\frac{R_{eff}-R_{1obs}\cos^2\theta}{R_{eff}+R_{ex}^{CEST}(\Delta\omega)/(1+f_m)+R_{ex}^{MT}(\Delta\omega)}}_{water\ saturaton} + \underbrace{\frac{R_{ex}^{CEST}(\Delta\omega)/(1+f_m)}{R_{eff}+R_{ex}^{CEST}(\Delta\omega)/(1+f_m)+R_{ex}^{MT}(\Delta\omega)}}_{CEST} + \underbrace{\frac{R_{ex}^{MT}(\Delta\omega)}{R_{eff}+R_{ex}^{CEST}(\Delta\omega)/(1+f_m)+R_{ex}^{MT}(\Delta\omega)}}_{MT} \tag{4}$$

This equation can be separated into three terms, each representing a specific effect. The first term represents the water saturation effect, the second term represents any CEST/NOE effects, and the third term represents the MT effect. Since the MT line shape is broader than the CEST/NOE line shape, the second term is modeled as a Lorentzian function. Similarly, the third term is also modeled as a Lorentzian function. Each term in Eq. (4) has different line width and offset, allowing them to be isolated through the multiple-pool Lorentzian fit.

**1)** Multiple-pool Lorentzian fitted $R_{eff}$

When the multiple-pool Lorentzian fitted water saturation effect (the first item in Eq. (4)) is set to zero, the signal S becomes the reference signal for quantifying the water saturation effect ($S_{ref_w}$). Eq. (4) can be rewritten as,

$$1 - \frac{S_{ref_w}(\Delta\omega)}{S_0} = \frac{R_{ex}^{CEST}(\Delta\omega)/(1+f_m)}{R_{eff}+R_{ex}^{CEST}(\Delta\omega)/(1+f_m)+R_{ex}^{MT}(\Delta\omega)} + \frac{R_{ex}^{MT}(\Delta\omega)}{R_{eff}+R_{ex}^{CEST}(\Delta\omega)/(1+f_m)+R_{ex}^{MT}(\Delta\omega)} \quad (5)$$

$S_{ref_w}$ can be derived from Eq. (5) as:

$$\frac{S_{ref_w}(\Delta\omega)}{S_0} = \frac{R_{eff}}{R_{eff}+R_{ex}^{CEST}(\Delta\omega)/(1+f_m)+R_{ex}^{MT}(\Delta\omega)} \quad (6)$$

Thus, $R_{eff}$ can be obtained from the ratio of Eq. (3) and Eq. (6),

$$R_{eff} = \frac{S_{ref_w}(\Delta\omega)}{S_{lab}(\Delta\omega)} R_{1obs} \cos^2 \theta \quad (7)$$

**2)** Multiple-pool Lorentzian fitted $R_{ex}^{MT}$

Since MT line shape is broader than that of CEST/NOE, most signals on the Z-spectrum used to fit the MT effect are far from water ($\Delta\omega > \omega_1$) and are not at the frequency offsets of CEST/NOE. Therefore, in this case, $R_{ex}^{CEST}(\Delta\omega) = 0$ and $R_{eff} \approx R_{1obs}$. Then, the actual fitted item for MT, denoted as $L_6(\Delta\omega)$, can be derived from Eq. (5),

$$L_6(\Delta\omega) = \frac{R_{ex}^{MT}(\Delta\omega)}{R_{1obs}+R_{ex}^{MT}(\Delta\omega)} \quad (8)$$

Subsequently, $R_{ex}^{MT}$ can be then obtained from Eq. (8):

$$R_{ex}^{MT}(\Delta\omega) = \frac{L_6(\Delta\omega)R_{eff}}{(1-L_6(\Delta\omega))} \approx \frac{L_6(\Delta\omega)R_{1obs}}{(1-L_6(\Delta\omega))} \quad (9)$$

**Supporting information Table S1.** Starting points and boundaries of the amplitude, width, and offset of all pools in the Lorentzian fit. The unit of peak width and offset is ppm.

| | Start | Lower | Upper |
|---|---|---|---|
| $A_{water}$ | 0.9 | 0.02 | 1 |
| $W_{water}$ | 1.4 | 0.1 | 10 |
| $\Delta_{water}$ | 0 | -1 | 1 |
| $A_{amide}$ | 0.025 | 0 | 0.2 |
| $W_{amide}$ | 0.5 | 0.4 | 3 |
| $\Delta_{amide}$ | 3.5 | 3 | 4 |
| $A_{amine}$ | 0.01 | 0 | 0.2 |
| $W_{amine}$ | 1.5 | 0.5 | 5 |
| $\Delta_{amine}$ | 2 | 1 | 3 |
| $A_{NOE(-1.6)}$ | 0.001 | 0 | 0.2 |
| $W_{NOE(-1.6)}$ | 1 | 0 | 1.5 |
| $\Delta_{NOE(-1.6)}$ | -1.5 | -2 | -1 |
| $A_{NOE(-3.5)}$ | 0.02 | 0 | 1 |
| $W_{NOE(-3.5)}$ | 3 | 1 | 5 |
| $\Delta_{NOE(-3.5)}$ | -3.5 | -4.5 | -2.5 |
| $A_{semi\text{-}solid\ MT}$ | 0.1 | 0 | 1 |
| $W_{semi\text{-}solid\ MT}$ | 25 | 10 | 100 |
| $\Delta_{semi\text{-}solid\ MT}$ | 0 | -4 | 4 |

**Supporting information Table S2.** Description of the parameters used.

| Parameter | Description |
|---|---|
| $f_m$ | Concentration of the MT pool |
| $f_s$ | Concentration of solute pool |
| $k_{sw}$ | Solute-water exchange rate |
| $L_6$ | Lorentzian fit for MT pool (Supporting Information Theory) |
| $r_{MT}$ | Scaling factor for MT pool |
| $r_{amines}$ | Scaling factor for amine pool |
| $R_{1w}$ | Water longitudinal relaxation rate |
| $R_{eff}$ | Effective water relaxation in rotating frame |
| $R_{ex}(\Delta\omega)$ | Exchange dependent CEST effect in rotating frame |
| $R_{ex}^{APT}(\Delta\omega)$ | APT effect in rotating frame |
| $R_{ex}^{MT}(\Delta\omega)$ | MT effect in rotating frame |
| $R_{ex}^{NOE}(\Delta\omega)$ | NOE effect in rotating frame |
| $R_{ex}^{amines}(\Delta\omega)$ | Amine CEST effect in rotating frame |
| $S_0$ | Control signals |
| $S_{lab}$ | Label signal |
| $S_{ref}$ | Reference signal |

**Supporting information Table S3.** List of all sample parameters used to create the partially synthetic CEST data using Eq. (3) and Eq. (5). Since the NOE(-1.6) effect is far from the APT effect, it was not included in creating the synthetic training data.

| | water | amide | NOE(-3.5) | amines | MT |
|---|---|---|---|---|---|
| $f_s$ (%) | 100 | 0.06:0.02:0.14 | 0.4:0.3:1.6 | - | - |
| $k_{sw}$ ($s^{-1}$) | - | 40:30:160 | 20 | - | - |
| $T_1$ (s) | 1.6:0.2:2.4 | 1.5 | 1.5 | - | - |
| $T_2$ (ms) | 40:20:120 | 2:1:5 | 0.5 | - | - |
| Δ (ppm) | 0 | 3.6 | -3.3 | - | - |
| r | - | - | - | 0.5:0.25:1.5 | 0.4:0.3:1.6 |

**Supporting information Table S4.** List of all sample parameters used to create the tissue-mimicking data using numerical simulation of the Bloch-McConnell equation.

| | water | amide | NOE(-3.5) | guanidine | amine | MT |
|---|---|---|---|---|---|---|
| $f_s$ (%) | 100 | 0.05:0.04:0.13 | 0.2:0.6:1.4 | 0.01:0.02:0.05 | 0.15:0.15:0.45 | 4:4:12 |
| $k_{sw}$ ($s^{-1}$) | - | 20:60:140 | 20 | 300:200:700 | 3000:2000:7000 | 25 |
| $T_1$ (s) | 1.5:0.4:2.3 | 1.5 | 1.5 | 1.5 | 1.5 | 1.5 |
| $T_2$ (ms) | 30:40:110 | 1:1.5:4 | 0.5 | 15 | 15 | 0.03:0.02:0.07 |
| Δ (ppm) | 0 | 3.6 | -3.3 | 2 | 3 | -2.3 |

**Supporting information Table S5.** List of all sample parameters used to create the fully synthetic CEST data for type 1 and type 2 simulations. For type 1 simulations, the $f_s$ of guanidine is fixed at 0.03%.

| | water | amide | NOE(-3.5) | guanidine | amine | MT |
|---|---|---|---|---|---|---|
| $f_s$ (%) | 100 | 0.06:0.02:0.14 | 0.4:0.3:1.6 | 0.01:0.02:0.05 | 0.2:0.1:0.6 | 4:4:16 |
| $k_{sw}$ ($s^{-1}$) | - | 40:30:160 | 20 | 500 | 5000 | 25 |
| $T_1$ (s) | 1.6:0.2:2.4 | 1.5 | 1.5 | 1.5 | 1.5 | 1.5 |
| $T_2$ (ms) | 40:20:120 | 1:2:5 | 0.5 | 15 | 15 | 0.05 |
| $\Delta$ (ppm) | 0 | 3.6 | -3.3 | 2 | 3 | -2.3 |

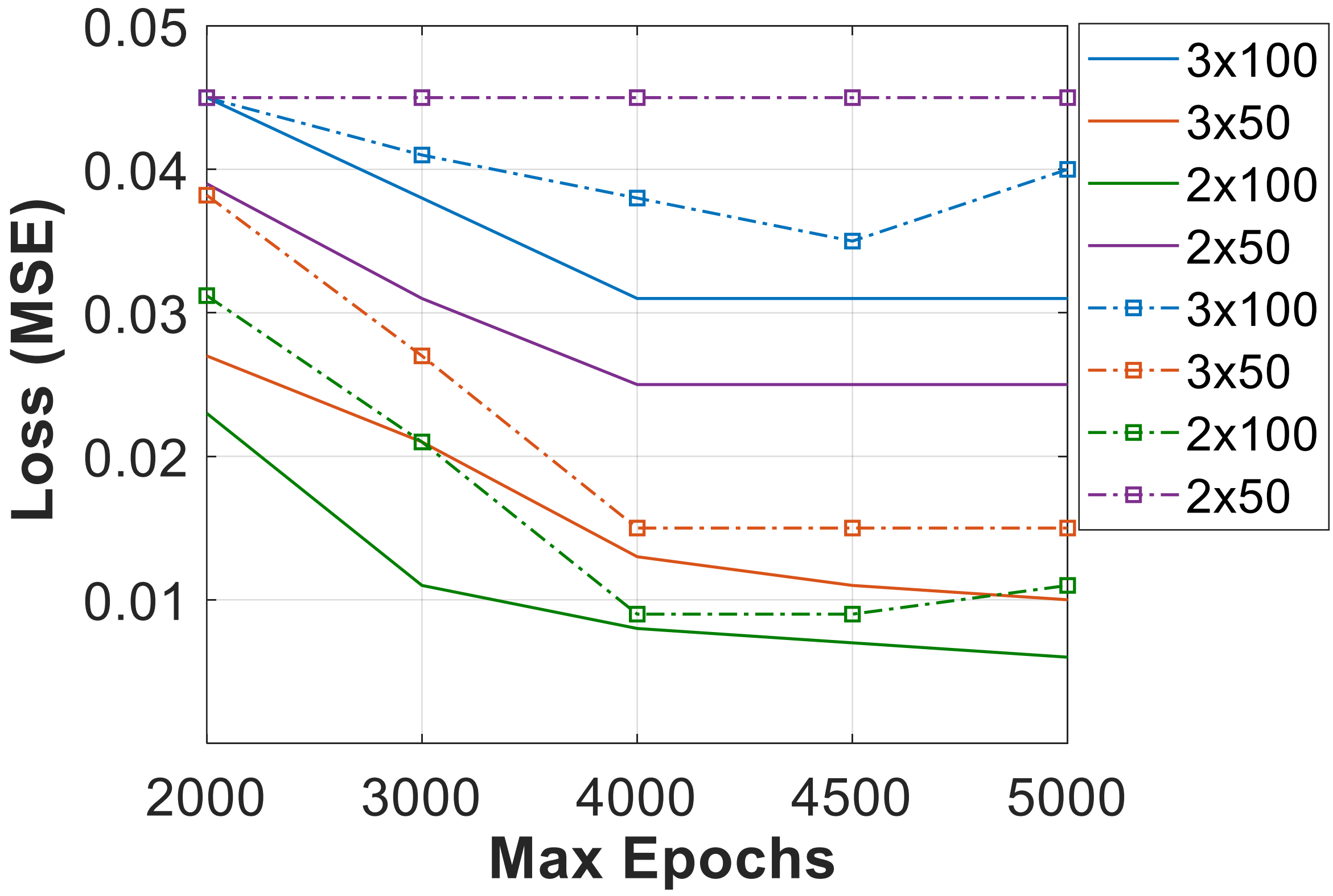

**Supporting information Fig. S1** Plot showing the mean squared error (MSE) loss versus the number of epochs for the training and validation using different model configurations. The solid line shows the training loss, while the dashed lines indicate the validation loss. The optimal training was found with 2 layers and 100 neurons with 4000 epochs. The learning rate was kept constant at $1\times10^{-3}$ with Adam optimizer and ReLU activation was used.

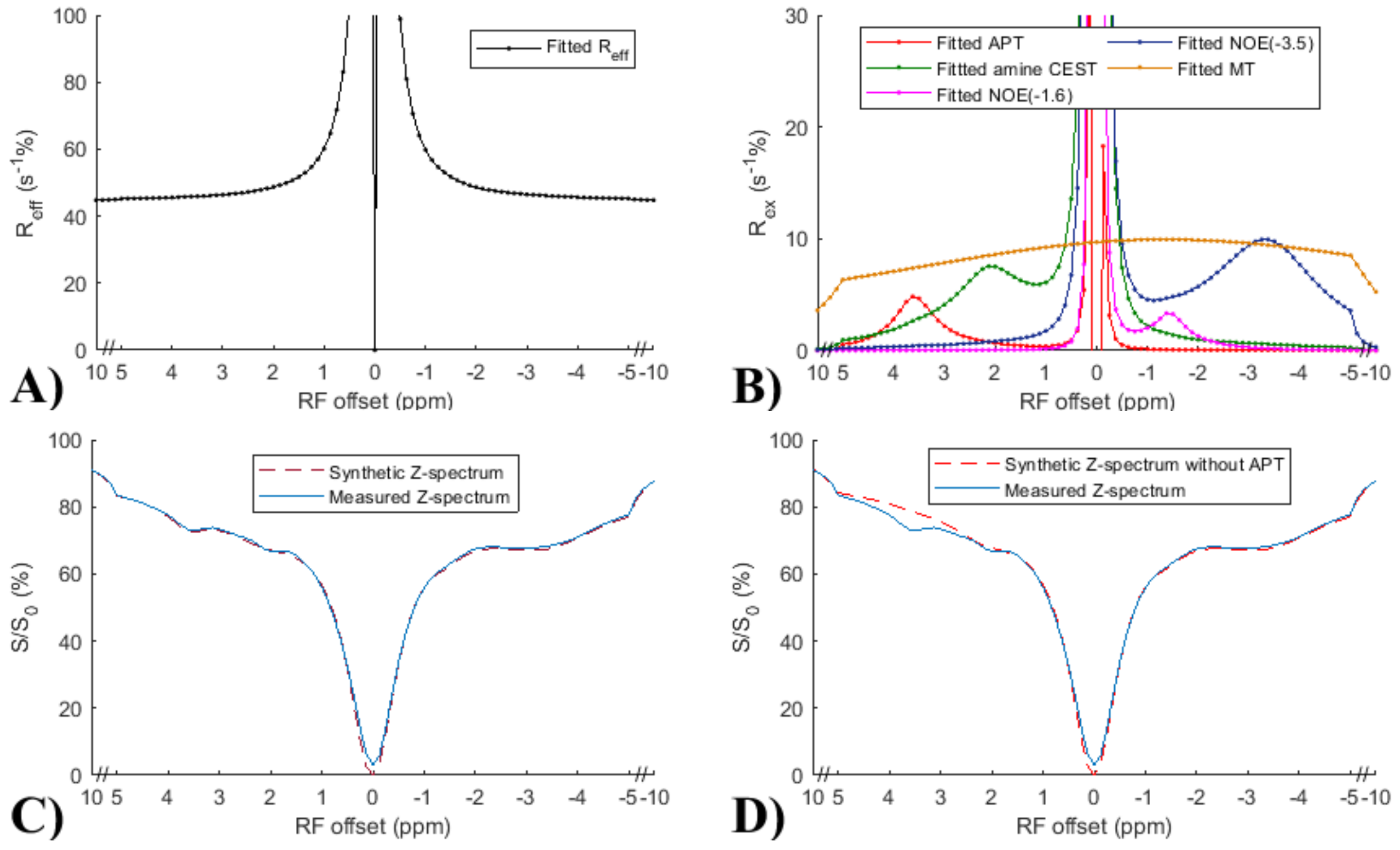

**Supporting information Fig. S2** Multiple-pool Lorentzian fitted $R_{eff}$ (a) and fitted $R_{ex}^{APT}$, $R_{ex}^{amines}$, $R_{ex}^{NOE(-1.6)}$, $R_{ex}^{NOE(-3.5)}$, $R_{ex}^{MT}$ spectra (b) from the average of the measured CEST Z-spectra in tumors in eight rat brains. Comparison between the measured CEST Z-spectra and synthetic Z-spectrum generated using all multiple-pool Lorentzian fitted components (c) as well as between the measured CEST Z-spectra and synthetic CEST Z-spectrum without APT (d).

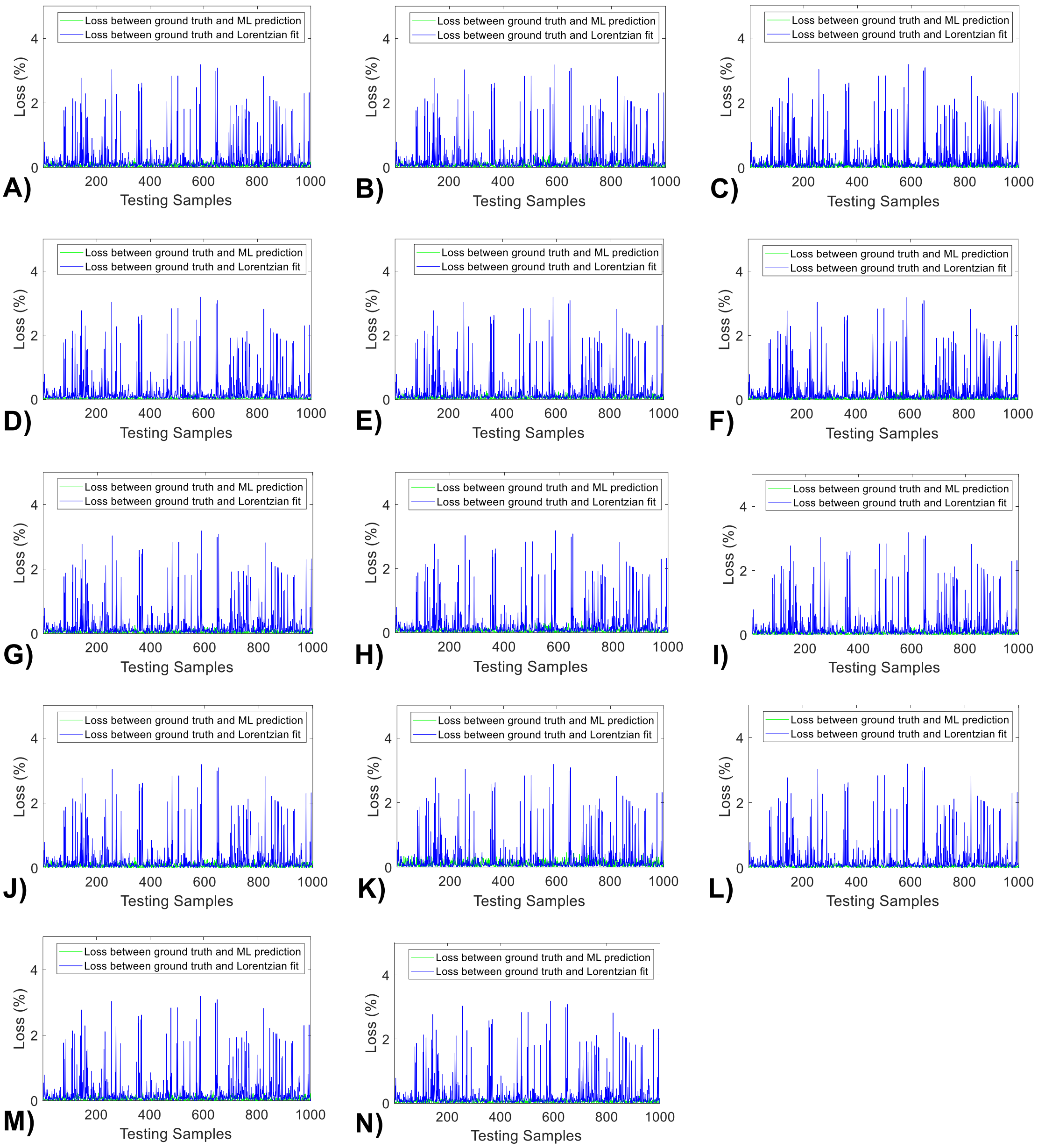

**Supporting information Fig. S3** Comparison of losses between the ML method and the multiple-pool Lorentzian fit for all testing data. ML models were trained on fourteen partially synthetic datasets, with the measured components fitted from ten other randomly selected Z-spectra (a-j), as well as the average of 50 (k), 100 (l), 500 (m), and 1000 (n) randomly selected Z-spectra, within the tissue-mimicking data. The mean losses are $7.2148\times10^{-4}$, $7.2766\times10^{-4}$, $7.2757\times10^{-4}$, $7.0909\times10^{-4}$, and $7.0852\times10^{-4}$ in (a-j), (k), (l), (m), and (n), respectively.

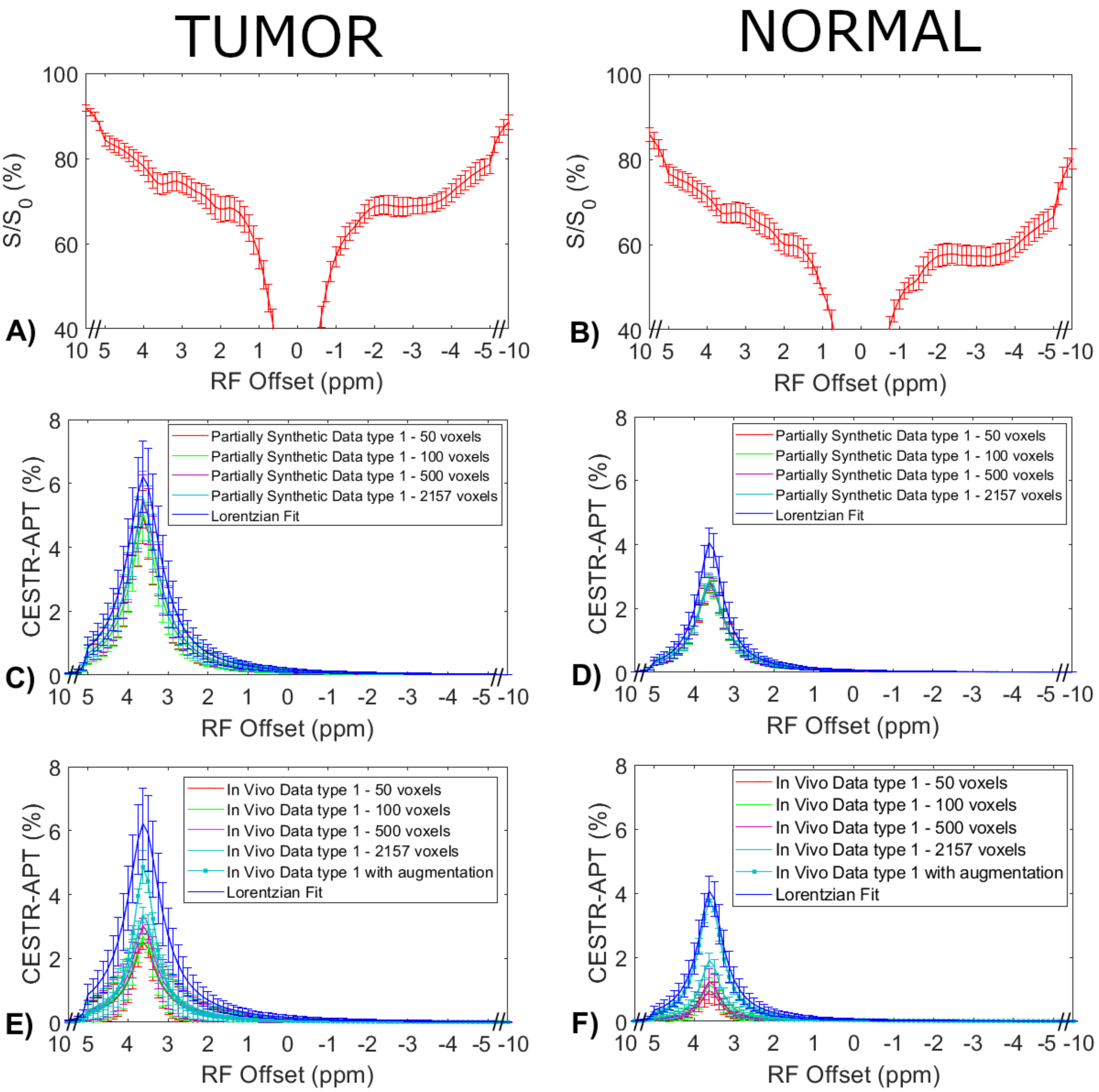

**Supporting information Fig. S4** Measured CEST Z-spectra from tumors and contralateral normal tissues in the three testing rats (a, b), the corresponding APT spectra from the ML prediction using the partially synthetic data for type 1 selection of 50, 100, 500, and 2157 voxels within five rat brains (c, d), and the corresponding APT spectra from the ML prediction using the measured in vivo data for type 1 selection of 50, 100, 500, and 2157 voxels within five rat brains as well as with data augmentation (e, f). The multiple-pool Lorentzian fitted APT spectra were also plotted in (c-f) for comparison.

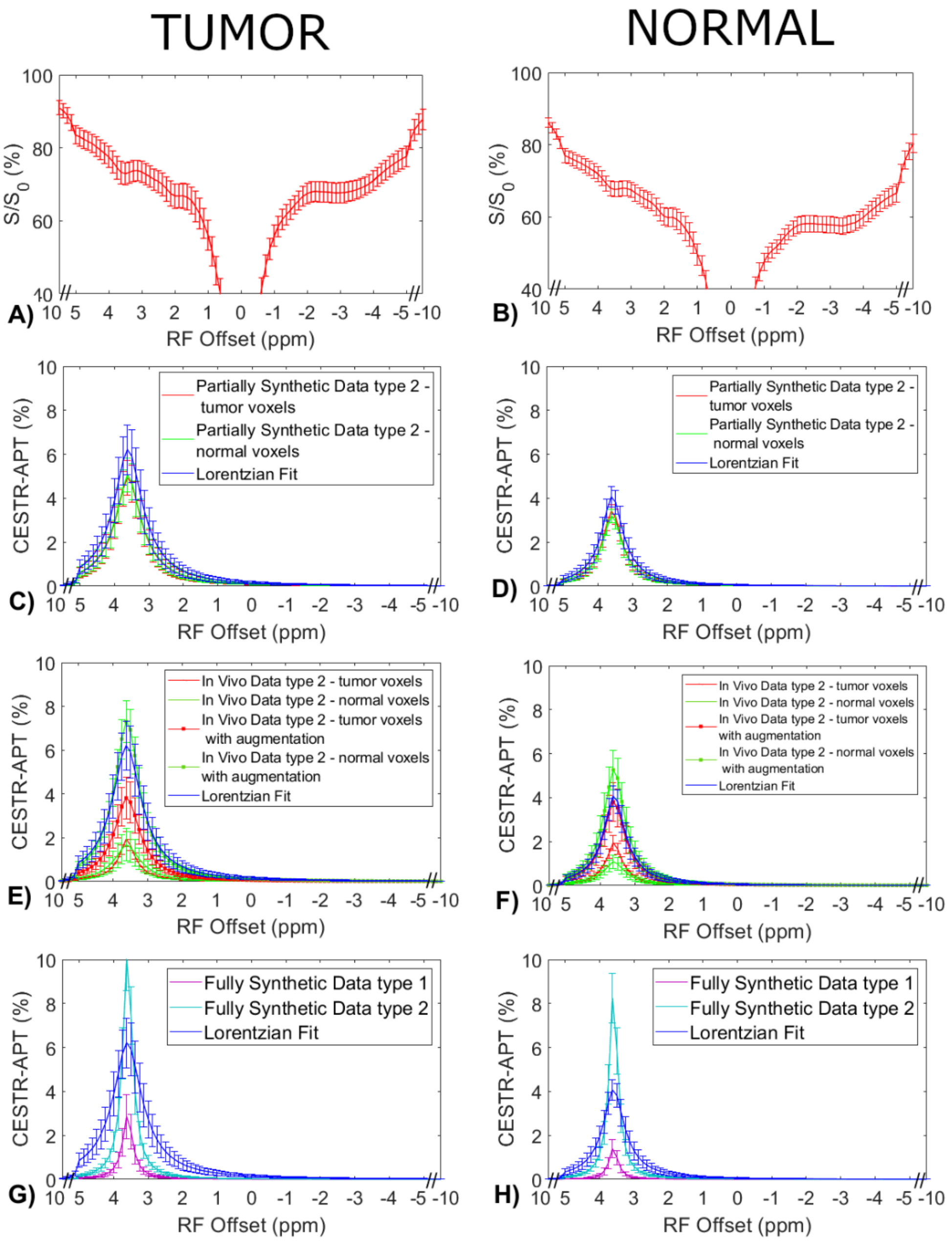

**Supporting information Fig. S5** Measured CEST Z-spectra from tumors and contralateral normal tissues in the eight rats (a, b), the corresponding APT spectra from the ML prediction using the partially synthetic data for type 2 selection of tumors and normal tissues (c, d), the corresponding APT spectra from the ML prediction using the measured in vivo data for type 2 selection of tumors and normal tissues (e, f), and ML prediction using fully synthetic data with the type 1 and type 2 simulations (g, h). The multiple-pool Lorentzian fitted APT spectra were also plotted in (c-h) for comparison.

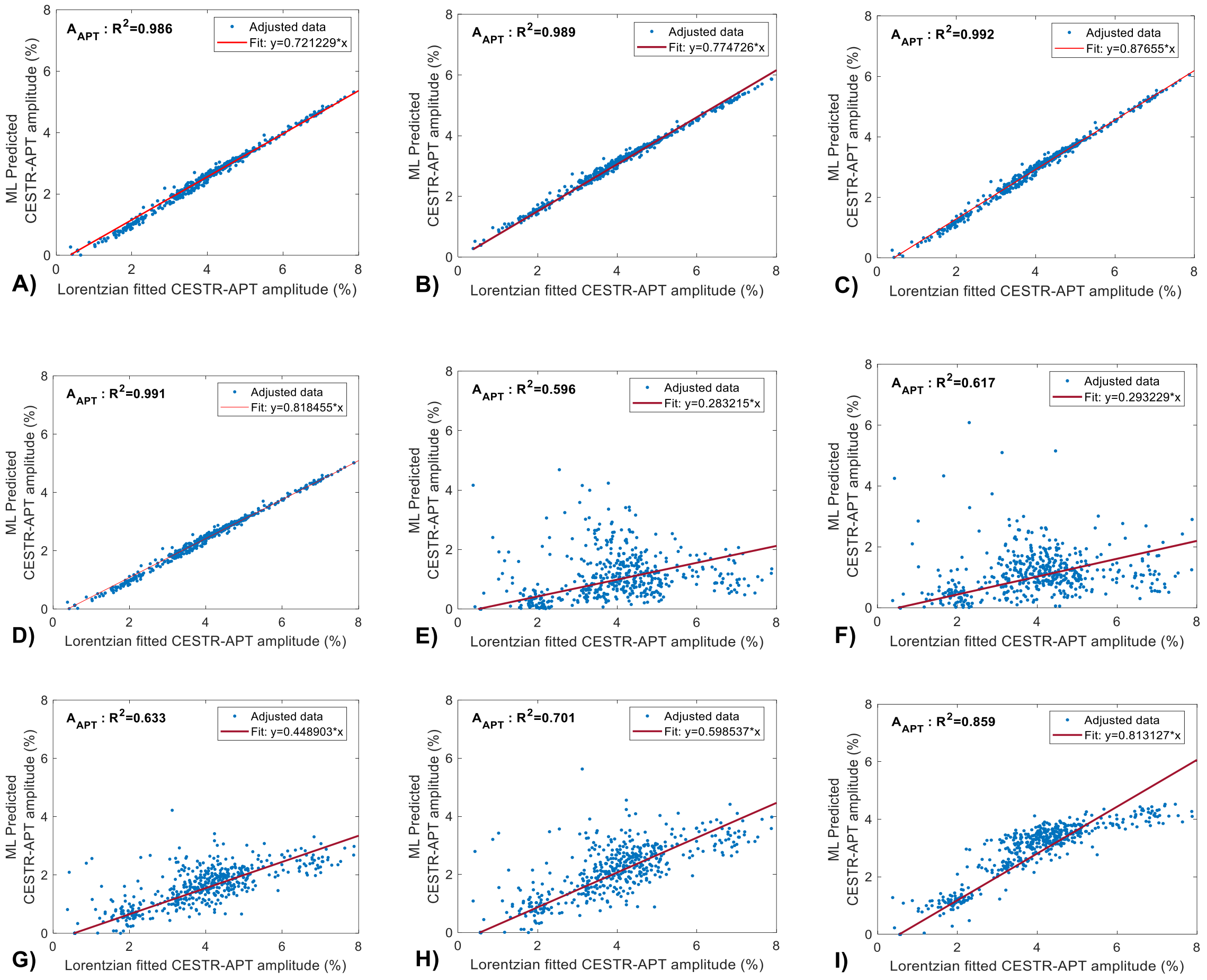

**Supporting information Fig. S6** Pixel-wise regression plots between the multiple-pool Lorentzian fitted APT amplitude and the predicted APT amplitude from all the ML models trained using the partially synthetic data for type 1 selection of 50 (a), 100 (b), 500 (c), and 2157 (d) voxels within five rat brains, using the measured in vivo data for type 1 selection of 50 (e), 100 (f), 500 (g), and 2157 (h) voxels within five rat brains as well as with data augmentation (i).

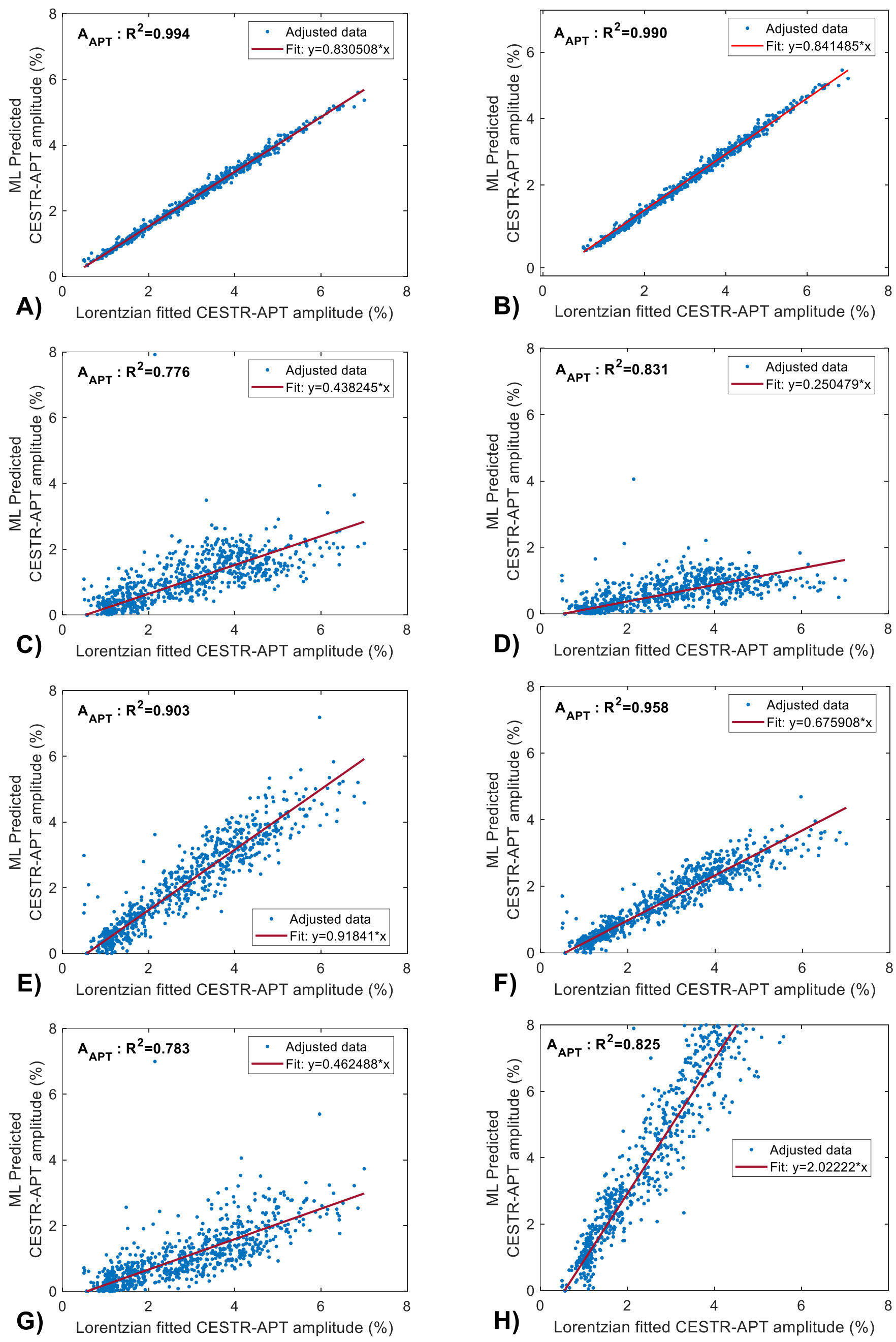

**Supporting information Fig. S7** Pixel-wise regression plots between the multiple-pool Lorentzian fitted APT amplitude and the predicted APT amplitude from all the ML models trained using the partially synthetic data for type 2 selection of tumors and normal tissues (a, b), using the measured in vivo data for type 2 selection of tumors and normal tissues (c, d), using the measured in vivo data for type 2 selection of tumors with data augmentation and normal tissues with data augmentation (e, f), and using fully synthetic data with the type 1 and type 2 simulations (g, h).

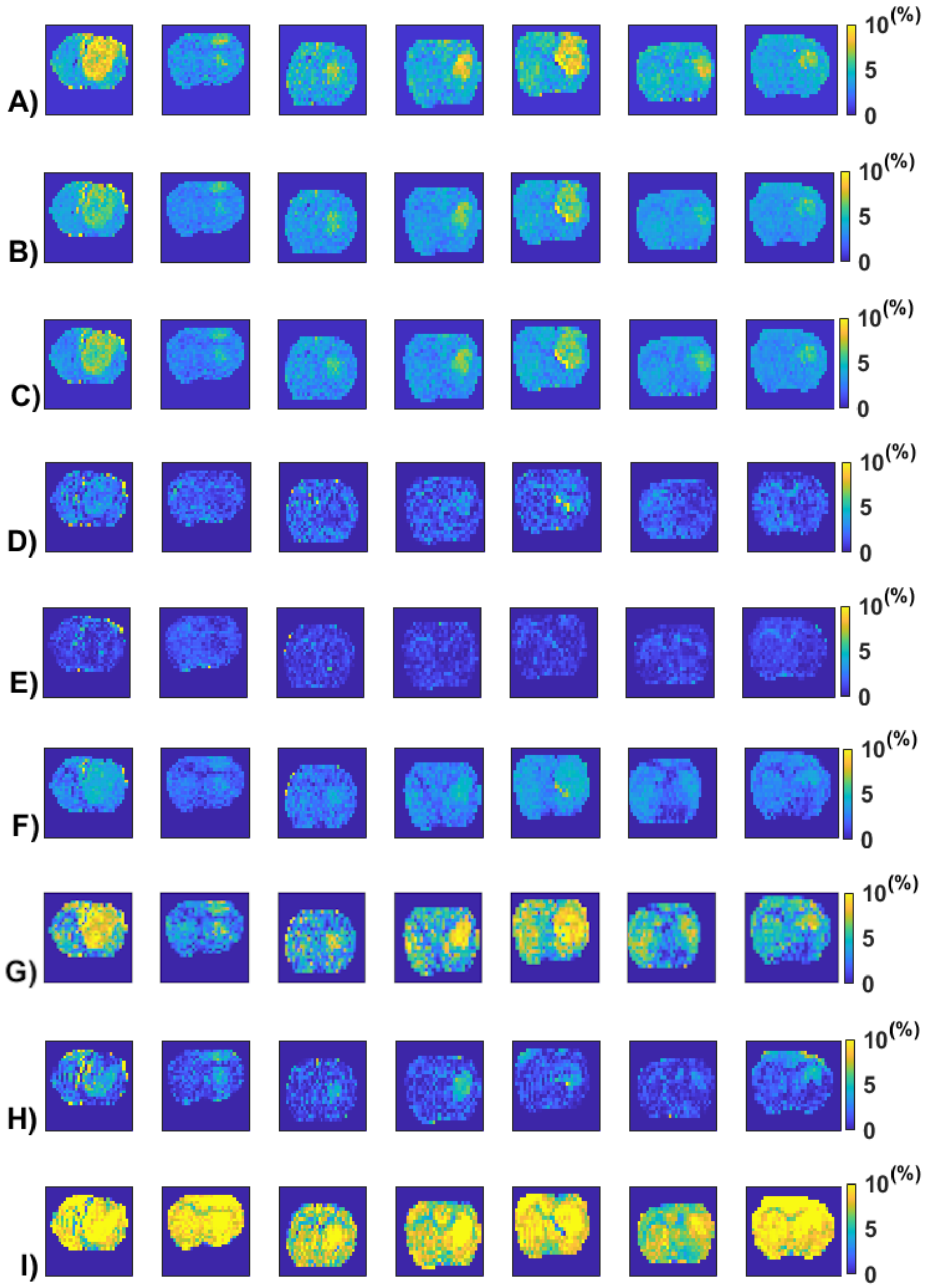

**Supporting information Fig. S8** APT amplitude maps from the seven rat brains (from left to right, Rat 2, Rat 3, Rat 4, Rat 5, Rat 6. Rat 7, Rat 8) using the Lorentzian fitting (a), ML prediction using partially synthetic data with the measured components from the in vivo data with type 2 selection of tumors (b) and normal tissues (c), ML prediction using the measured in vivo data with type 2 selection of tumors (d) and normal tissues (e), ML prediction using the measured in vivo data with type 2 selection of tumors with data augmentation (f) and normal tissues with data augmentation (g), as well as ML prediction using fully synthetic data with the type 1 (h) and type 2 (i) simulations.

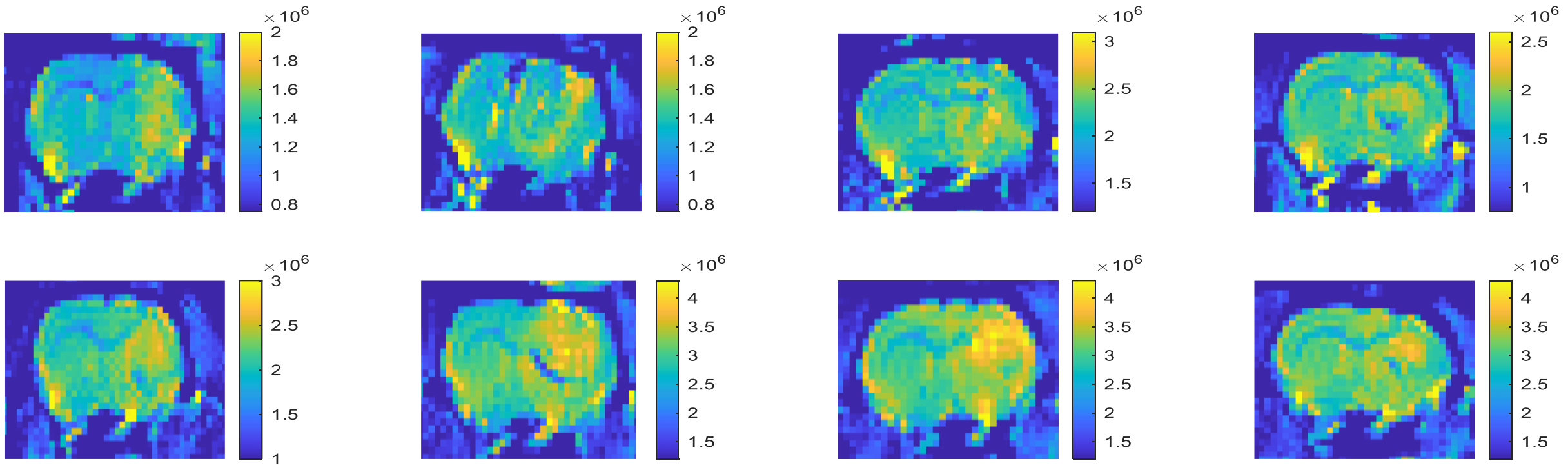

**Supporting information Fig. S9** $T_2$-weighted anatomy images from the eight rat brains. From left to right, Row 1 shows anatomy images for Rat 1 to Rat 4 respectively, and Row 2 shows the anatomy images for Rat 5 to Rat 8 respectively.

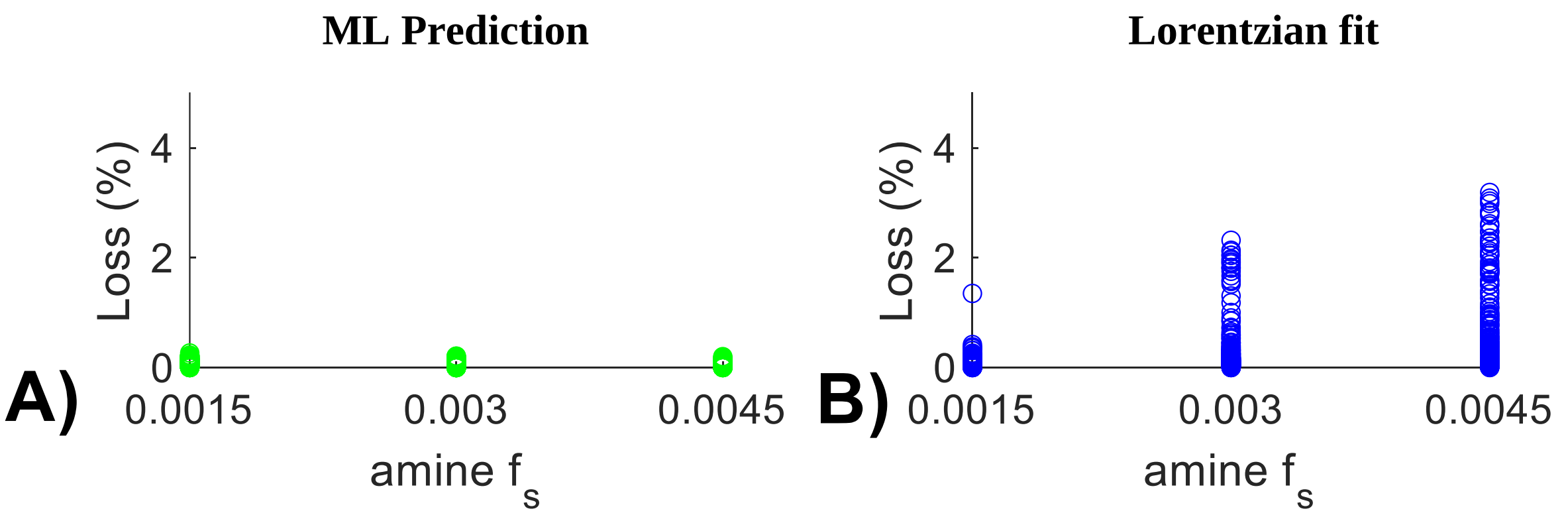

**Supporting information Fig. S10** Scatter plots of losses from the ML method using the partially synthetic data (a) and from the multiple-pool Lorentzian fit (b) of all testing tissue-mimicking data vs. amine $f_s$. The partially synthetic data were obtained with the measured components fitted from a randomly selected Z-spectrum within the tissue-mimicking data.

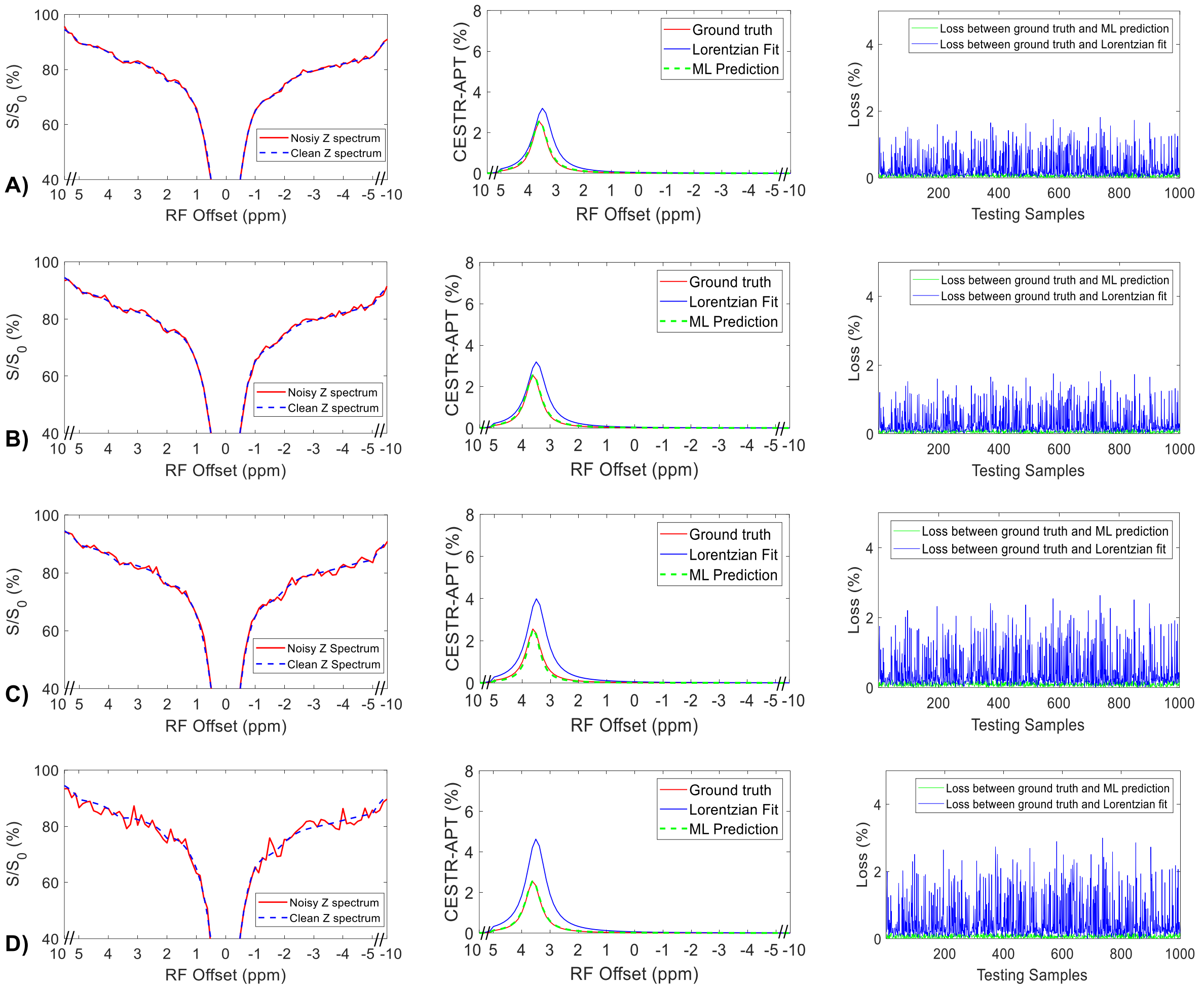

**Supporting information Fig. S11** Left column: a representative Z-spectrum from the tissue-mimicking data (clean) together with a few noisy Z-spectra with SNR of 200 (a), 150 (b), 100 (c), 50 (d). Middle column: A comparison of the corresponding APT spectra from the ML prediction, multiple-pool Lorentzian fit, and ground truth. Right column: A comparison of the corresponding losses between the ML prediction and the multiple-pool Lorentzian fit for all testing data.